\documentclass[preprint]{aastex6}

\usepackage{graphicx}
\usepackage{amssymb}
\usepackage{amsmath}
\usepackage{enumitem}
\usepackage{natbib}
\usepackage{booktabs}
\usepackage{multirow}

\shorttitle{GMC Clump Catalog: Method and Initial Results}
\shortauthors{Zetterlund et al.}
\slugcomment{\today}     

\begin{document}
\title{A Galactic Molecular Cloud Clump Catalog from Hi-GAL Data: Method and Initial Results Comparison to BGPS} 

\author{Erika Zetterlund, Jason Glenn}
\affil{CASA, Department of Astrophysical and Planetary Sciences, University of Colorado 389-UCB, Boulder, CO 80309, USA}
\email{erika.zetterlund@colorado.edu}
\and
\author{Erik Rosolowsky}
\affil{Deptartment of Physics, University of Alberta, Edmonton, Alberta, Canada}


\begin{abstract}

As the precursors to stellar clusters, it is imperative that we understand the distribution and physical properties of dense molecular gas clouds and clumps. Such a study has been done with the ground-based Bolocam Galactic Plane Survey (BGPS). Now the \textit{Herschel} infrared GALactic plane survey (Hi-GAL) allows us to do the same with higher quality data and complete coverage of the Galactic plane.
We have made a pilot study comparing dense molecular gas clumps identified in the Hi-GAL and BGPS surveys, using six $2^\circ \times 2^\circ$ regions centered at Galactic longitudes of $\ell = 11^\circ$, 30$^\circ$, 41$^\circ$, 50$^\circ$, 202$^\circ$, and $217^\circ$. We adopted the BGPS methodology for identifying clumps and estimating distances, leading to 6198 clumps being identified in our substudy, with 995 of those having well-constrained distances. These objects were evenly distributed with Galactic longitude, a consequence of Hi-GAL being source confusion limited. These clumps range in mass from 10$^{-2}$ M$_\odot$ to 10$^{5}$ M$_\odot$, and have heliocentric distances of up to 16 kpc. When clumps found in both surveys are compared, we see that distances agree within 1 kpc and ratios of masses are of order unity. This serves as an external validation for BGPS and instills confidence as we move forward to cataloging the clumps from the entirety of Hi-GAL. In addition to the sources that were in common with BGPS, Hi-GAL found many additional sources, primarily due to the lack of atmospheric noise. We expect Hi-GAL to yield $2\times10^5$ clumps, with 20\% having well-constrained distances, an order of magnitude above what was found in BGPS.

\end{abstract}

\section{Introduction}

Significant observational and theoretical progress has been made in the study of star formation. However, crucial aspects, such as why the stellar and star cluster initial mass functions appear uniform across many Galactic environments, remain unexplained. In order to understand stellar clusters and OB associations, it is critical to understand that from which they are formed. Thus the study of molecular cloud clumps has become a primary focus in the field of high-mass star formation \citep[e.g.,][]{McKee07}. Large-scale studies of the Galactic dense molecular gas --- its distribution and properties --- are necessary in order to challenge models of galaxy evolution and reveal the origin of the stellar cluster initial mass function.

Such large-scale studies have recently become practical with the actualization of Galactic Plane dust continuum surveys at sub/millimeter wavelengths [BGPS: \citet{Aguirre2011,BGPS9}; ATLASGAL: \citet{Schuller09,Csengeri14,Csengeri16}; Hi-GAL: \citet{Molinari2010,Molinari16b}]. These surveys have detected tens of thousands of molecular cloud clumps and cores, which can now be extracted and used to study physical properties of high-mass star formation regions. The census of Galactic dense molecular cloud structures enabled by these surveys will constrain star formation and galaxy evolution theories \citep[e.g.,][]{Kennicutt12}. Such studies of physical properties \citep[e.g.,][]{Peretto10,Giannetti13,Elia13} and mass distributions \citep[e.g.,][]{Netterfield09,Olmi13,Olmi14,Gomez14,BGPS12,Wienen15} have begun, but have yet to reveal a coherent story concerning the evolution of the dense interstellar medium and the uniformity of the stellar cluster mass function.

Fortunately, theoretical models of the dense interstellar medium are beginning to produce predictions robust enough to be used in conjunction with observational data to constrain the molecular cloud clump mass function \citep[e.g.,][]{Donkov12,Veltchev13}. This function is generally expected to take either a power-law or lognormal form, with each distribution corresponding to a different physical process in the molecular clouds. Gravitational collapse of dense structures would produce a power-law \citep[e.g.,][]{Padoan02}. A lognormal density distribution can be produced by supersonic turbulence \citep[e.g.,][]{Padoan97}, although such conditions are not necessary \citep{Tassis10}. Likely, these processes are interacting within the dense molecular clouds, producing a mass function displaying a combination of these forms \citep[e.g.,][]{Offner14,Hopkins13}. Which mode dominates will have implications for the competing theories of high-mass star formation \citep{Elmegreen85}.

On a grander scale, the distribution of dense molecular gas in the Milky Way has implications for using our Galaxy as the ground truth for studies of galaxy evolution. Until recently, numerical models had substantial difficulty generating long-lived spiral galaxies. This was due either to rapid depletion of gas before the disks could be sustained or the galaxies blowing themselves apart with an inadequate balance of gravity and stellar feedback. In the past two decades, remarkable progress has been made towards reproducing observed properties of galaxies \citep[e.g.,][]{Katz96,Keres09,Stinson13}. Currently, simulations that incorporate feedback into the interstellar medium from star formation and the late stages of massive stellar evolution can reliably create large spiral galaxies. However, the star formation, being unresolved in the simulations, uses prescriptive recipes tuned a posteriori. Typically, star formation is turned on at a fixed interstellar medium density threshold (e.g. $10^3$ cm$^{-3}$), at which point $\sim1\%$ of the gas mass is converted into stars \citep{Kim2014}. Unfortunately, numerical simulations have largely been unable to a priori produce the strong winds and inefficiency of star formation observed in galaxies \citep{Hopkins14}. The dense molecular gas distribution is an observational key to probing the star formation efficiency on scales resolvable by simulations. Because the amount and rate of star formation drives feedback, it is imperative that these recipes are tested against reality.

We will use the \textit{Herschel Space Observatory's} Hi-GAL survey, which provides complete coverage of the Galactic Plane in the far-infrared to submillimeter wavebands, to create such a map and test these explanations of clump mass functions and galaxy evolution models. This paper describes the method we will be using to identify Hi-GAL molecular cloud clumps and determine their heliocentric distances. It will furthermore compare the results of a number of representative test regions to BGPS, a well-vetted benchmark. However, we expect Hi-GAL to significantly supersede BGPS. \citet{Molinari16b} have independently begun cataloging the dense molecular cloud clumps found in the Hi-GAL survey. They have identified and extracted photometry for clumps found in the majority of the inner Galaxy in all five of \textit{Herschel's} photometric bands. This work serves as a complement to the work of \citet{Molinari16b} through the use of a fundamentally different source identification technique. In addition, we go beyond their work in the determination of distances and physical properties of the clumps.

Dense molecular gas structures can be divided into three categories: clouds, clumps, and cores. Molecular clouds have denser substructures called molecular cloud clumps, which in turn have even denser substructures called molecular cloud cores. These cores are gravitationally bound and will form individual stars or simple stellar systems. Typical radii are clouds $R = 1-7.5$ pc, clumps $R = 0.15-1.5$ pc, and cores $R = 0.015-0.1$ pc. Typical densities are clouds $50-500$ cm$^{-3}$, clumps $10^3-10^4$ cm$^{-3}$, and cores $10^4-10^5$ cm$^{-3}$ \citep{Bergin2007}. Cores are only resolvable within a couple kiloparsecs by single-dish telescopes. Clumps are detectable within $\sim 7$ kpc, beyond which the resolution is no longer sufficient for \textit{Herschel}. Farther out we can only resolve entire clouds \citep{BGPS7}.

\section{Data}

Previously, the Bolocam Galactic Plane Survey (BGPS) was made with the Caltech Submillimeter Observatory and was released in two versions \citep{Aguirre2011,BGPS9}. BGPS was a $\lambda = 1.1$ mm survey over the longitudes $-10^\circ < \ell < 90^\circ$ and latitudes $|b| < 0.5^\circ$, plus an additional 20 deg$^2$. From the survey, a catalog of 8,594 (Version 2) sources was produced, 20\% of which have well constrained distances. Even with its incomplete coverage of the Galactic Plane, BGPS showed the emergence of spiral arms, as well as evidence for significant amounts of dense interarm gas \citep{BGPS12}.

The \textit{Herschel} infrared GALactic plane survey (Hi-GAL) \citep{Molinari2010} originally planned to observe a $2^\circ$ strip covering $| \ell | < 60^\circ$, but was extended to include the entire $360^\circ$ of the Galactic plane. Using the SPIRE \citep{Griffin2010} and PACS \citep{Poglitsch2010} instruments aboard the \textit{Herschel Space Observatory} (HSO), this survey observed in wavebands at 70, 160, 250, 350, and 500 $\mu$m. SPIRE (250, 350, and 500 $\mu$m) observed thermal dust emission, which is concentrated in dense molecular gas clumps, whereas PACS (70 and 160 $\mu$m) observed dense gas and extended warm structures to a greater extent than the longer-wavelength SPIRE bands. Owing to the low optical depth of thermal dust in the submillimeter continuum, Hi-GAL allows us to view dense molecular gas throughout the entire Galactic disk. The flux density-limited gas mass sensitivity is 250 M$_\odot$ ($5\sigma$) at distances up to 20 kpc (corresponding with a $1\sigma$ RMS of 100 mJy at 250 $\mu$m, assuming a temperature of 20 K).

In this paper, we analyzed a representative sample of six Hi-GAL regions, centered at $\ell = 11^\circ$, 30$^\circ$, 41$^\circ$, 50$^\circ$, 202$^\circ$, and $217^\circ$. This sample includes regions from  both inner and outer portions of the Galaxy, and, with the exception of $\ell=202^\circ$, have a significant overlap with BGPS observations. Figure 1 shows the maps for the $\ell = 30^\circ$ and 217$^\circ$ regions. Emission levels vary strongly depending on proximity to the Galactic center, but there is extensive structure in all regions.

\begin{figure}[!h]
\begin{center}
	\includegraphics[width=0.49\textwidth]{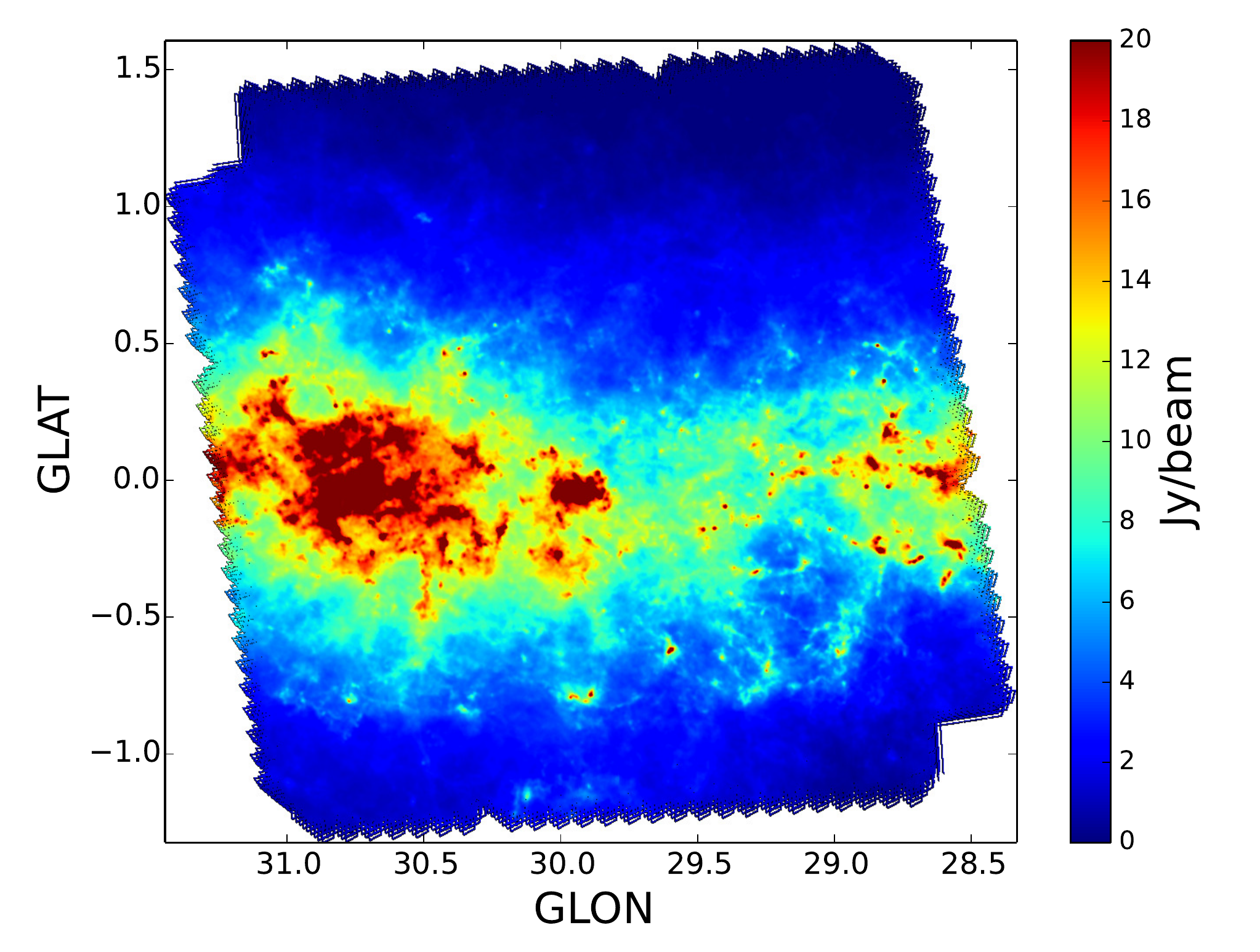}
	\includegraphics[width=0.49\textwidth]{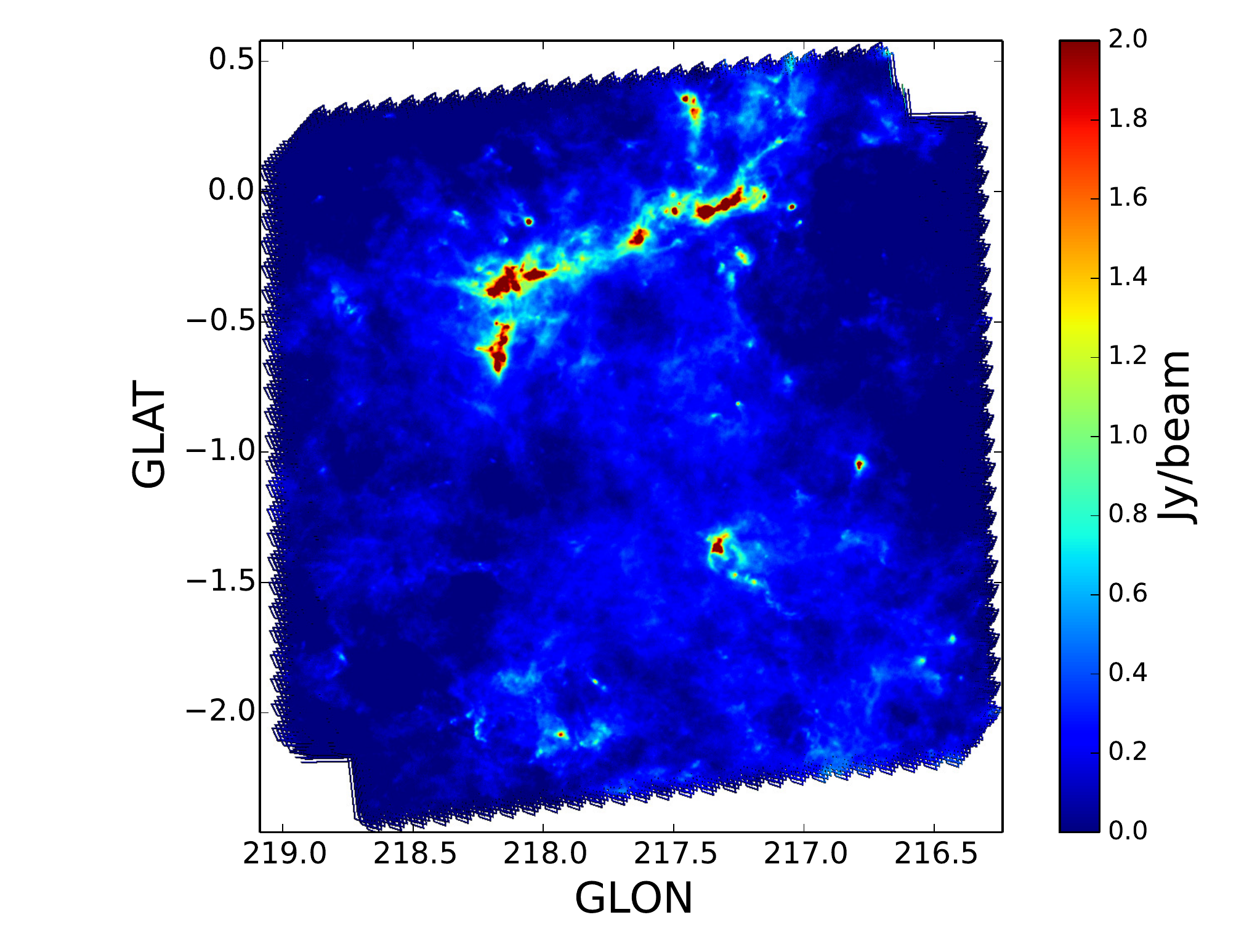}
\caption{Example maps in SPIRE 500 $\mu$m from our representative sample of Hi-GAL --- $\ell = 30^\circ$ (\textit{left}), which looks through the edge of the Galactic bar, and $\ell=217^\circ$ (\textit{right}). Emission levels are significantly higher in the inner Galaxy than the outer Galaxy due both to the greater quantities of molecular gas, and to the longer optical path integrating more emission.}
\label{fig:exmaps}
\end{center}
\end{figure}

\section{Method}

\subsection{Map Making}
The Hi-GAL observations are stored in the Herschel Science Archive (HSA) and the Herschel Interactive Processing Environment (HIPE) allows users to access the data.\footnote{\url{http://www.cosmos.esa.int/web/herschel/hipe-download}} The tool UniHIPE was used to format the HSA data suitable to be input into the Hi-GAL mapping pipeline, Unimap \citep{Piazzo2012,Piazzo2015a,Piazzo2015b}.

Each Hi-GAL observation is composed of two orthogonal scans. Unimap takes the time-ordered bolometer data from each scan and combines them to make a single astrophysical map, along with various evaluation maps. The pipeline begins by cleaning the time-ordered data. It removes the offset in the input data, then finds and removes signal jumps and glitches, both of which can be caused by cosmic rays. Finally, the drift, a low frequency signal due to slowly varying bolometer temperatures, is fit and removed. The time-ordered data are then made into a map, using an iterated Generalized Least Squares method. This method often leads to cross-like artifacts on bright sources, which are removed by Unimap's post-processing.

\subsection{Filtering}
The 500 $\mu$m SPIRE band has the lowest resolution of the bands, with a beam FWHM of $32\arcsec.2$, and so we convolved and rebinned all maps to 500 $\mu$m resolution to enable multiband photometry. Foreground cirrus clouds are present in all of the maps. Since this undesirable flux density is present on angular scales larger than that of molecular cloud clumps, the maps were high-pass filtered on the $3\arcmin$ scale to remove this flux density. The filtering was done in frequency-space using an inverted 2D Gaussian window with $\sigma$ corresponding to $3\arcmin$. This scale was chosen for two reasons. Firs t, it is a larger angular scale than the largest of the BGPS clumps. Second, when the Hi-GAL maps were filtered on a larger scale of $\sigma = 5\arcmin$, nearly all of the clumps found were smaller than $3\arcmin$. Thus, this filtering scale removes as much extended emission as possible with minimal attenuation of molecular cloud clump sizes and flux densities (and therefore inferred masses).

Simulations were performed in order to quantify the effects of the high-pass filter. Synthetic Gaussian objects were inserted onto both the $\ell=30^\circ$ map, and onto constant background maps. Synthetic sources were distributed randomly, but were restricted to have their centers no closer than three times the source FWHM. The maps were then filtered and objects identified by Bolocat (See Section 3.3) on the constant background map. These source masks were then used to calculate source properties on synthetic sources from the $\ell=30^\circ$ background map. This was done for filtering scales of $\sigma=2.4\arcmin$, $3.2\arcmin$, $4.1\arcmin$, and $4.8\arcmin$, and with various synthetic source sizes. For each combination of filtering scale and source size $\sim$1000 sources were generated, using multiple maps where necessary.

Figure 2 shows median attenuation factors (ratios of post-filtering recovered quantities to input quantities) for integrated flux density ($S$), peak flux density ($S_\text{pk}$), and FWHM. Our mean clump size corresponds most closely to the second smallest synthetic object size ($1.2\arcmin$), and 90\% of our objects are smaller than the dashed line. Thus Hi-GAL clumps will have their integrated flux density attenuated by $(20\pm10)\%$ for our choice of a filtering scale of $\sigma=3\arcmin$. Similarly, peak flux density will be attenuated by $<1\%$, and FWHM will be attenuated by $(4\pm3)\%$.

\begin{figure}[!h]
\begin{center}
	\includegraphics[width=0.32\textwidth]{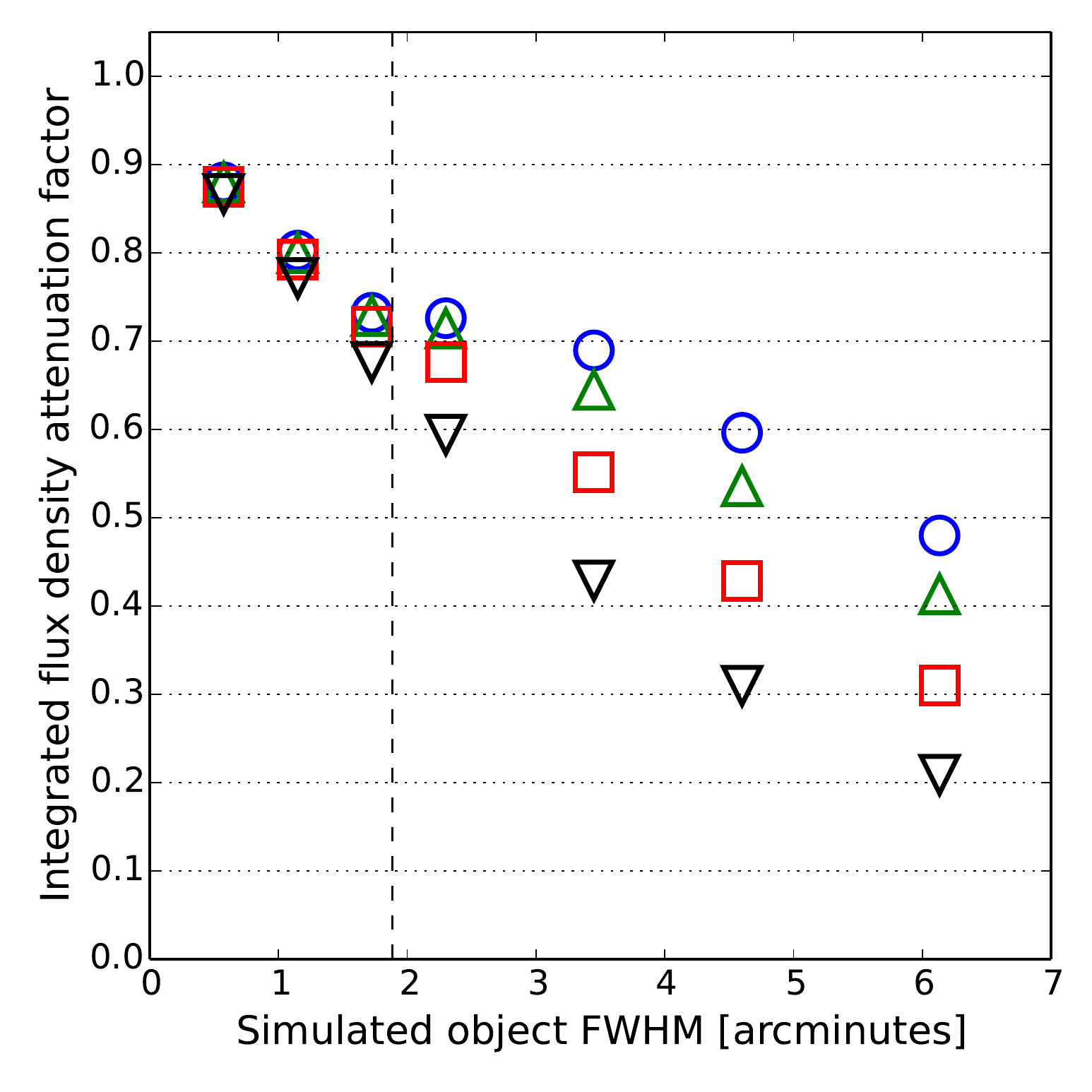}
	\includegraphics[width=0.32\textwidth]{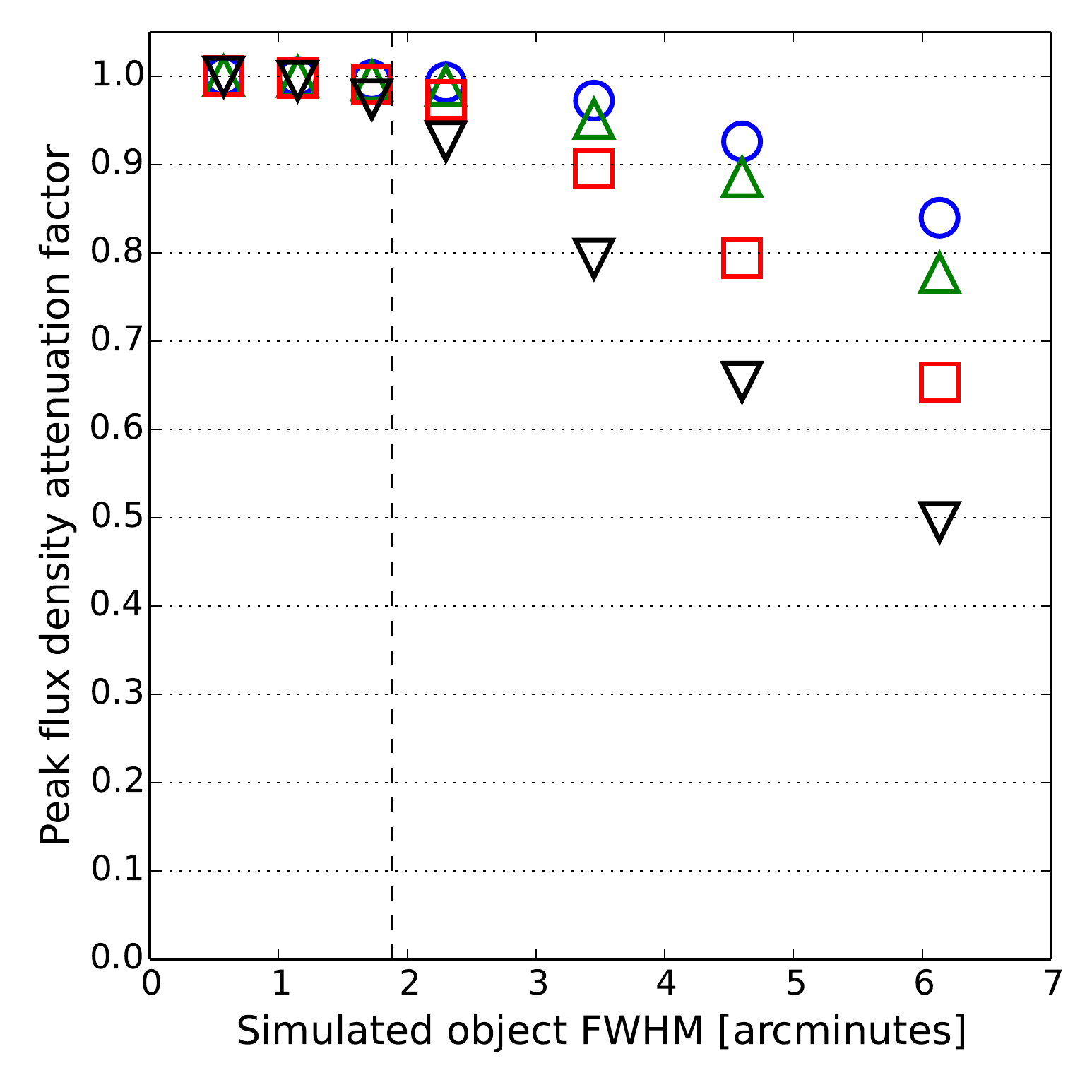}
	\includegraphics[width=0.32\textwidth]{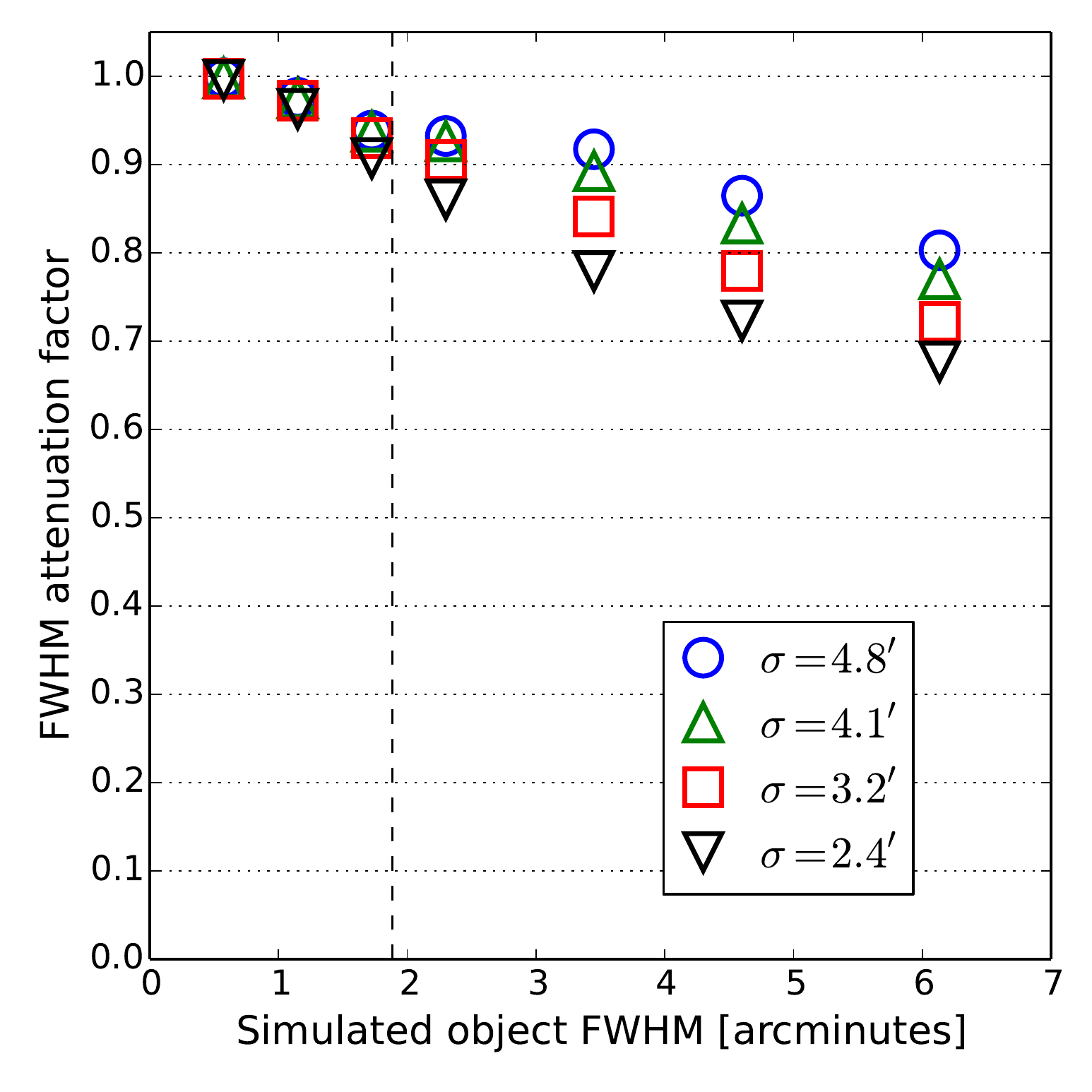}
\caption{Attenuation factors for integrated flux density (\textit{left}), peak flux density (\textit{center}), and FWHM (\textit{right}). Maps of simulated objects are filtered at various spatial scales. Plotted are the median attenuation factors of objects identified by Bolocat. The dashed line indicates the 90$^\text{th}$ percentile in angular size for objects in our Hi-GAL substudy.}
\label{fig:sims}
\end{center}
\end{figure}

While the majority of the observation regions are covered by both of \textit{Herschel's} orthogonal scans, the edges are not. We mask off the regions which are not covered by both scans to exclude these lower-quality data and to avoid confusing the clump-finding code with the original ragged map edges. This removed 29\% of flux-containing pixels. We also masked off regions where the high-pass filter added flux density, as opposed to removing it. This was a consequence of the flux density on large scales being negative in these areas, and thus subtracting those negative values added flux density to the pixels. These areas were located at high galactic latitude, where flux density was low. This removed an additional $3-7$\%, depending on the field, of flux-containing pixels. The masking process is demonstrated in Figure 3.

\begin{figure}[!h]
\begin{center}
	\includegraphics[width=0.32\textwidth]{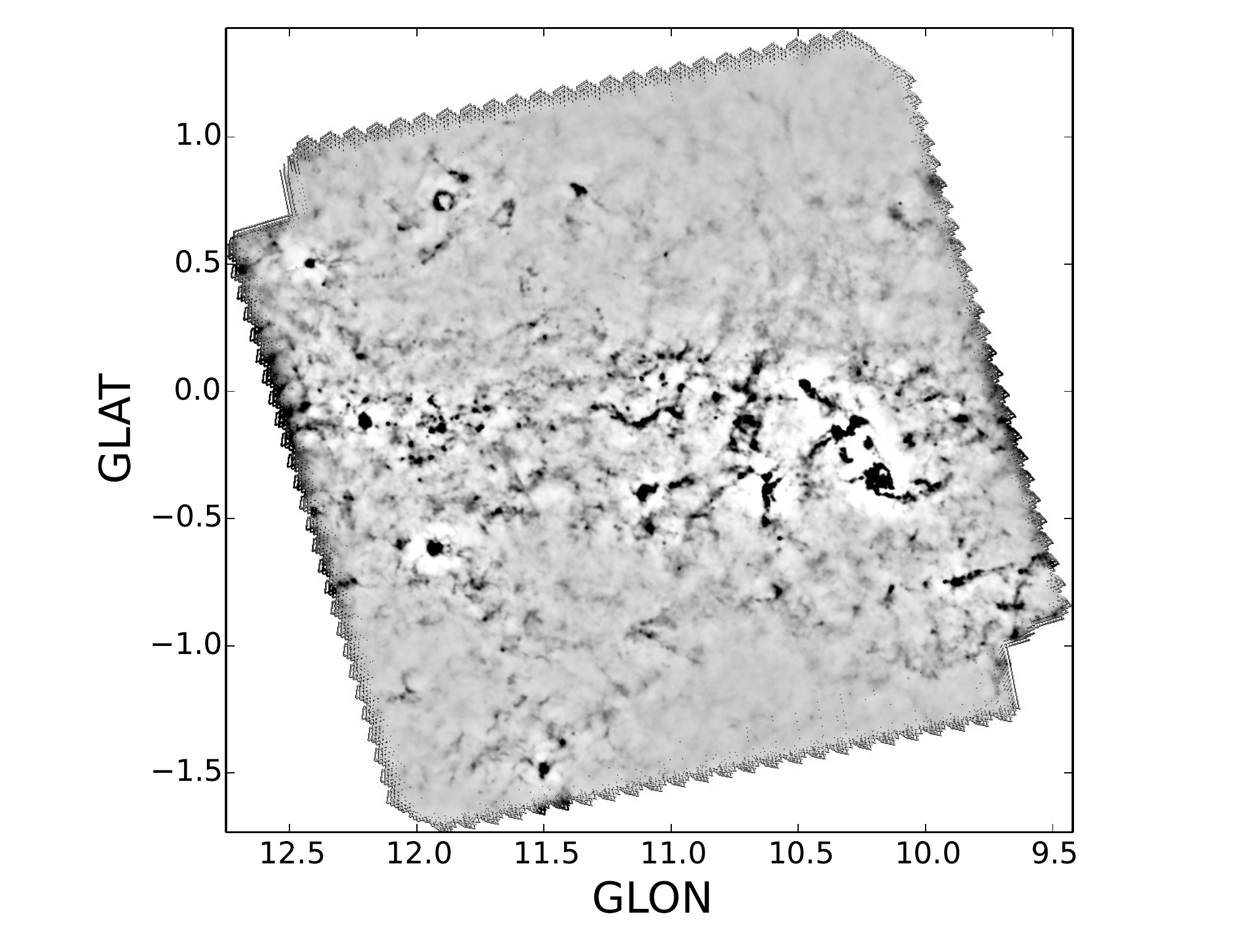}
	\includegraphics[width=0.32\textwidth]{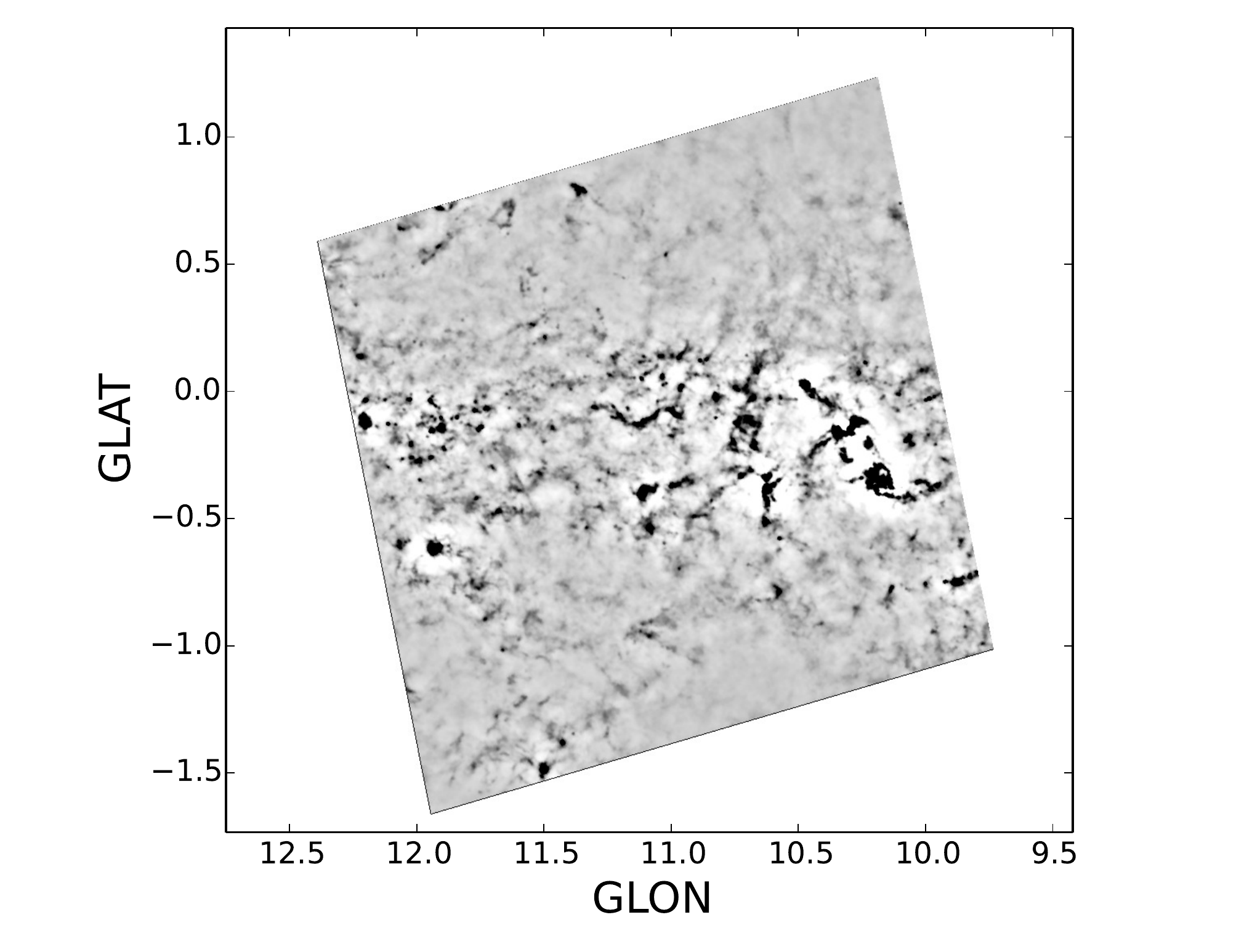}
	\includegraphics[width=0.32\textwidth]{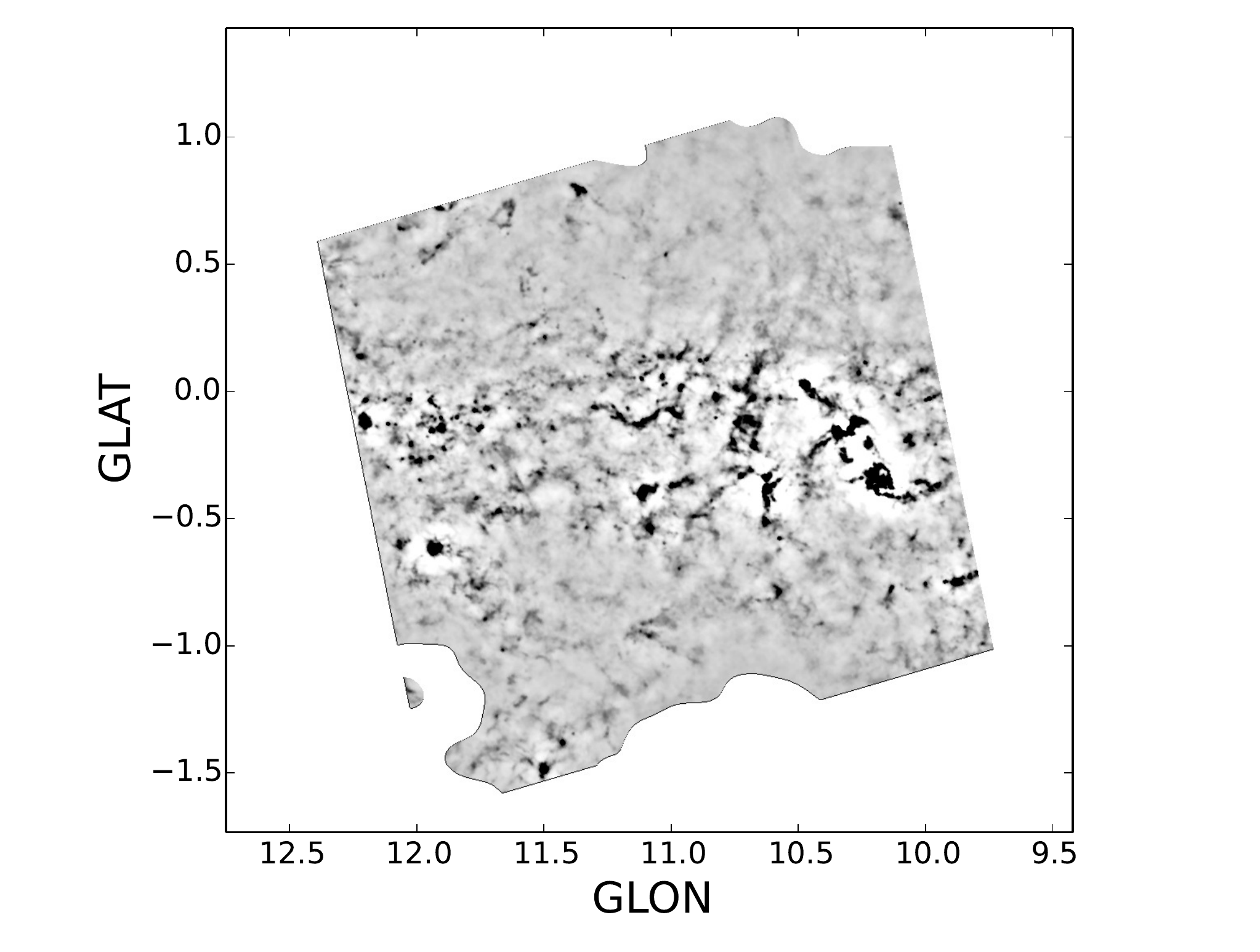}
\caption{Masking the $\ell=11^\circ$ region. \textit{Left}: The complete high-pass filtered map. The two scan directions can be seen. \textit{Center}: Pixels not covered by both scans have been masked off. \textit{Right}: Pixels to which the high-pass filter added flux density have been masked off.}
\label{fig:masking}
\end{center}
\end{figure}

\subsection{Source Identification}
Source identification was done on 500 $\mu$m SPIRE maps using Bolocat, a seeded watershed algorithm tool developed for BGPS \citep{BGPS2}. Clumps are found by first identifying regions of significant emission, where significance is determined in units of the local noise estimate. These regions of high significance are then expanded into adjacent lower significance areas, eliminating artificial small-scale structure. Finally, regions are split into multiple clumps, where appropriate, based on local maxima.

Hi-GAL is confusion limited by cirrus, large-scale structure, and source confusion that in BGPS were attenuated by an atmospheric subtraction algorithm. The flux density level of clumps which we can identify in each map is limited by the emission levels in the map. This is demonstrated in the power spectral densities (PSDs) in Figure 4. There is greater power along lines of sight with more interstellar medium (ISM), as well as greater power on larger scales, which may be due to source confusion or the sizes of GMCs themselves. A typical GMC with a radius of 10 pc \citep[e.g.][]{Solomon87} would have to be $>23$ kpc away in order to have a 3$\arcmin$ extent.

\begin{figure}[!h]
	\begin{center}
	\includegraphics[width=0.6\textwidth] {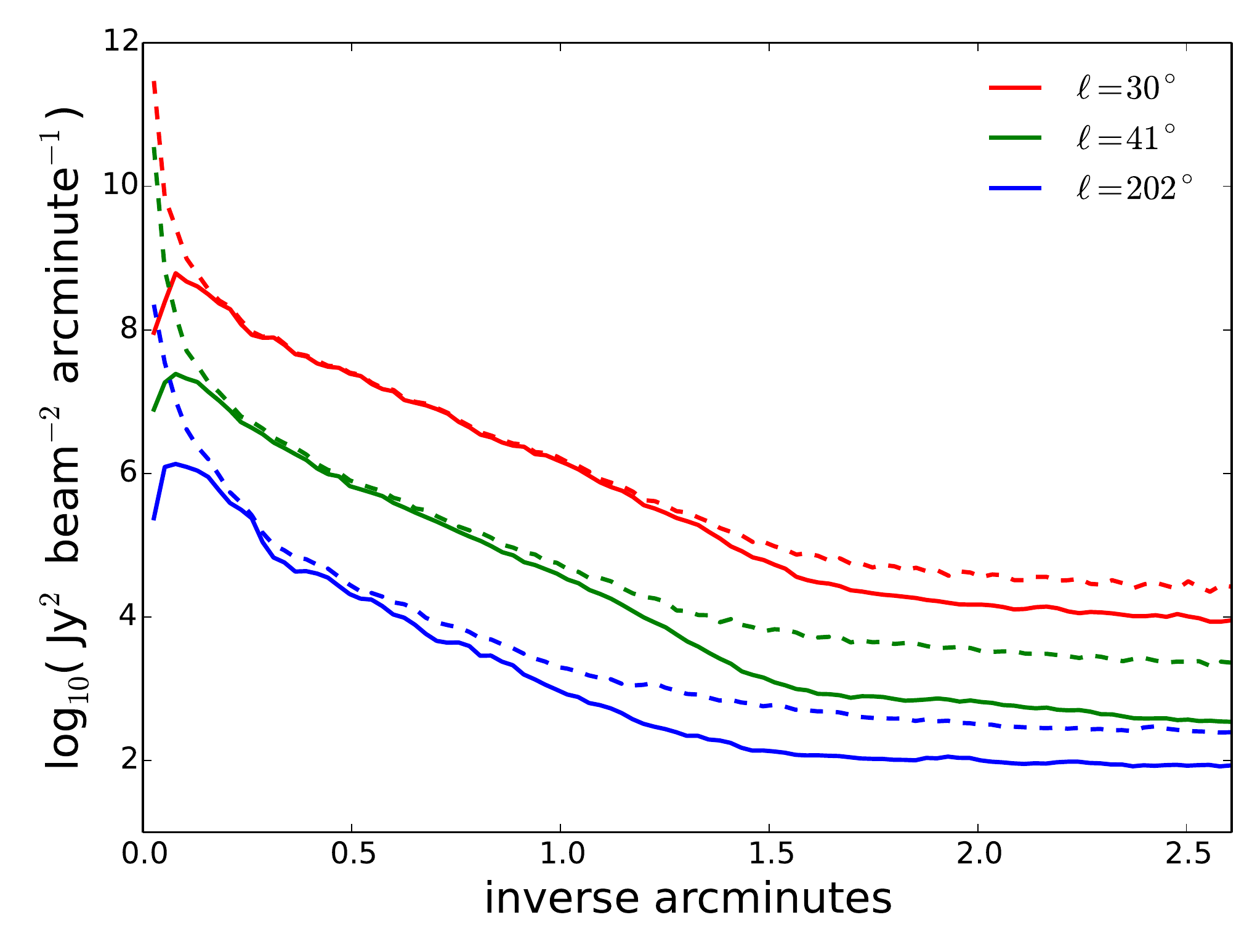}
	\end{center}
	\caption{The PSDs for a representative sample of regions. Pre- and post-filtering are shown as dashed and solid lines, respectively. Emission levels on all spatial scales decrease with distance from the Galactic center and lines of sight through the inner Galaxy have greater integrated flux density than towards the outer Galaxy.}
	\label{fig:psd_em}
\end{figure}

With more overall emission in regions nearer the Galactic center, sources in those regions must be brighter for us to detect them above the confusion noise. Therefore we used a constant-noise map for each region, with each ``noise" map's value determined by the emission level in its respective emission map. This was done by fitting the distribution flux densities of pixels with positive emission to an exponential function. The noise value was taken to be proportional to the scale factor of the exponential fit, specifically $0.25\lambda$, where $\lambda$ is the scale factor. As an objective choice was not possible, this selection was made as it consistently produced object contours which matched our visual expectations across regions at widely varying Galactic longitudes. The specific value of $0.25\lambda$ for the noise was chosen after an exploration of Bolocat's parameter space, which was done using the $\ell=41^\circ$ region as a test map.

The threshold for detection and the level down to which areas meeting the detection threshold were expanded were determined first. We chose to set the threshold at $3\sigma$ and to expand down to $1\sigma$, where $\sigma$ is the ``noise" level. These values were chosen as they successfully included the flux density which had the appearance of real structure. A division criterion of $2\sigma$ was then decided upon. That is, if the saddle between two local maxima is different from the maxima by at least $2\sigma$, then the clump is divided. This was chosen as it produced results consistent with what was seen in the flux density contours of the test map, a sample section of which can be seen in Figure 5. The initial threshold and division criterion were the same in BGPS, which only expanded down to $2\sigma$. 

\begin{figure}[!h]
	\begin{center}
	\includegraphics[width=0.75\textwidth] {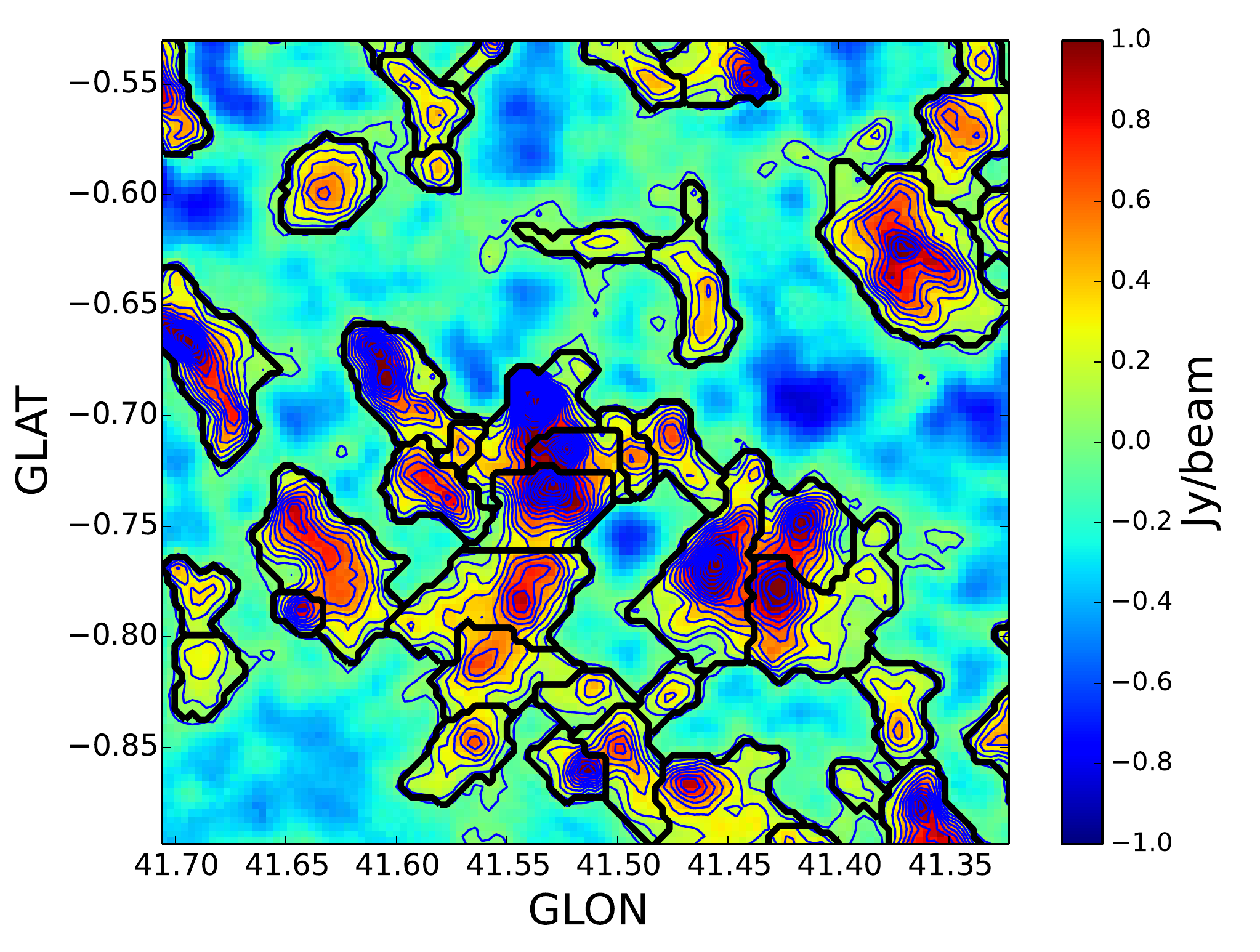}
	\end{center}
	\caption{Subsection of the test map for determining Bolocat parameters. Thick black contours show object borders. Thin blue contours show flux density levels with a step size of $2\sigma$, starting at zero.}
	\label{fig:testmap}
\end{figure}

Choosing a higher threshold would have decreased the number of sources found, as would have choosing a higher floor down to which to expand the regions, since clumps must be at least a beam in size. A higher expansion floor would have also decreased the size of clumps, particularly further from the mid-plane, where flux density is lower. Choosing a lower threshold for splitting clumps would also have led to smaller clumps, although clumps sizes would be limited by the requirement that local maxima must be separated by at least two beam widths. Our choices for these parameters were chosen to match our visual expectations in the test map, attempting to err on the side of a higher detection threshold and less clump division.

The high-pass filter also has associated systematic effects. A more aggressive filter would have resulted in smaller clumps. More significantly, the high-pass filter affects the flux densities of the identified clumps. Figure 6 compares high-pass filtered and unfiltered flux densities in the $\ell=50^\circ$ region. Clumps were identified in the filtered map, with photometry re-calculated using the unfiltered maps for the comparison. Photometry was done by summing the flux density found in the pixels within the source boundary, as opposed to background-subtracted aperture photometry. No clump exceeds a flux density ratio of unity by construction, as the small percentage of pixels where the filter added flux density were excluded from source finding.

There is a trend towards the clump containing a higher fraction of the total flux density along the line of sight for higher flux density clumps. On the other hand, the faintest clumps have a wide range of flux density ratios. These trends are seen in all of the regions in this substudy.

\begin{figure}[!h]
\begin{center}
	\includegraphics[width=0.6\textwidth]{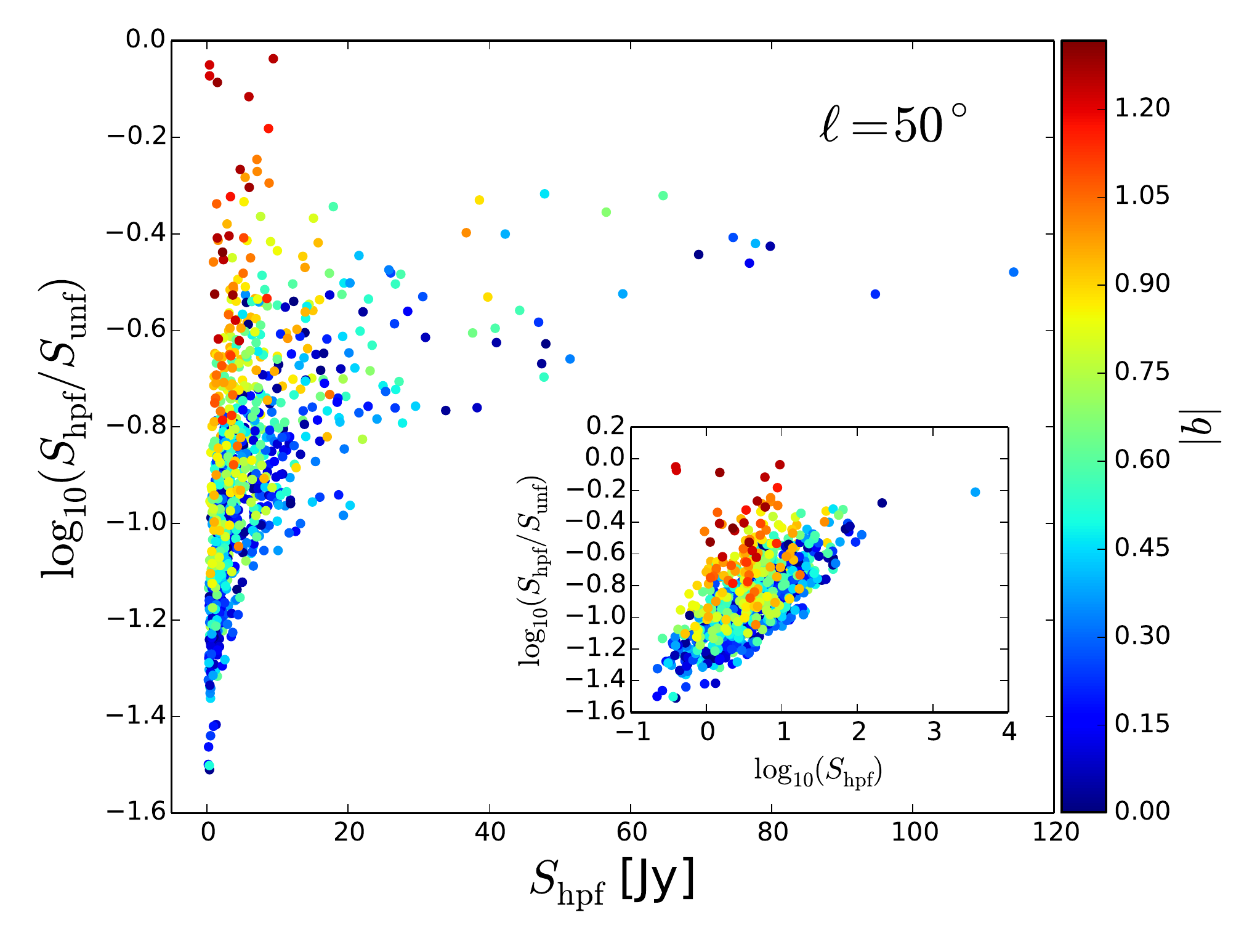}
\caption{Comparison of high-pass filtered and unfiltered flux densities for the $\ell=50^\circ$ region. $S_{\text{hpf}}$ is the clump's total flux density, taken from the high-pass filtered map in which it was identified, and $S_{\text{unf}}$ is the total flux density in the unfiltered map which lies within the clump's border. Colors correspond to absolute value of Galactic latitude.}
\label{fig:filterflux}
\end{center}
\end{figure}

\subsection{Distance Determination}

This distance determination method was developed for BGPS by \citet{BGPS8} and \citet{BGPS12}, and resulted in 1,710 well-constrained distances. Full details of the method are available in \citet{BGPS8,BGPS12}, and a summary is provided here. To construct a 3D map from the sources identified in Hi-GAL and to compute their physical properties, heliocentric distances must be determined. Every line of sight has a function $V_r(d_\odot)$, found from a Galactic rotation curve, describing radial velocity as a function of heliocentric distance. Kinematic distances can be determined by matching a clump's $v_\text{LSR}$ to $V_r(d_\odot)$. However, within the inner Galaxy, this information results in two possible kinematic distances and an ambiguity as to which is the true distance. This ambiguity will be broken using probabilistic methods in a Bayesian framework. A unique distance probability density function (DPDF) will be calculated for each source. The posterior DPDF is found by
\begin{equation}
	\text{DPDF} = \mathcal{L}(d_\odot | \ell,b,v_\text{LSR}) \prod_i P_i (d_\odot | \ell,b),
\end{equation}
where $\mathcal{L}(d_\odot | \ell,b,v_\text{LSR})$ is the kinematic distance likelihood, and the $P_i (d_\odot | \ell,b)$ are various Bayesian priors which will help determine which distance has the higher probability of being true. Once DPDFs are calculated, they can be drawn from in Monte Carlo simulations, even when distances are not well constrained, and thus cloud clump properties can be characterized, with robust uncertainties. 

The rotation curve from the \citet{Reid14} BeSSeL survey, derived from maser parallax measurements to sites of high-mass star forming regions ($\mathbf{-12^\circ \le \ell \le 240^\circ}$), was paired with line-of-sight velocities derived from the $^{13}\text{CO}$ Galactic Ring Survey (GRS) to determine the kinematic distances to the molecular cloud clumps. Because there are often multiple velocity components in GRS spectra, each source spectrum is created by subtracting the spectrum of an off-source region from that of the on-source region. The off-source region is determined by creating a rind around the source mask, then excluding pixels associated with other Hi-GAL sources. In this way we eliminate the velocity components not associated with the clump from the clump's spectrum \citep[see][Section 4]{BGPS12}. Dense gas tracer observations of specific sources, originally made for BGPS, were also utilized. These observations were associated with the new Hi-GAL sources through the object masks obtained from Bolocat.
The most powerful $P_i$ implemented in BGPS used absorption by the clumps of diffuse Galactic mid-infrared emission near $\lambda=8$ $\mu$m. When the majority of this diffuse emission lies in the background, the molecular cloud clump is called an infrared dark cloud (IRDCs) \citep{Perault96,Simon06,Peretto09,Battersby11}. \citet{BGPS8} defined the term 8 $\mu$m absorption feature to include molecular cloud clumps which exhibit any $\lambda = 8$ $\mu$m intensity decrement, thus allowing for absorption less pronounced than seen in IRDCs. The $P_i$ developed to take advantage of this feature uses the Galactic infrared emission model of \citet{Robitaille12} to simulate clumps at various heliocentric distances. These simulated images are then compared to the corresponding GLIMPSE \citep{Churchwell09} IRAC Band 4 image, using a $\chi^2$ statistic, to generate the $P_i$.

The EMAF morphological matching of synthetic GLIMPSE based on Hi-GAL flux density and actual GLIMPSE maps works far better with high-pass filtered maps than with unfiltered maps. In the unfiltered maps, the presence of cirrus emission contributes too much flux to the clumps, causing them to be placed much further away than is plausible. Use of a $\sigma=3\arcmin$ Gaussian high-pass filter removes the cirrus emission and remedies this problem. Two examples from the $\ell=41^\circ$ region are shown in Figure 7.

\begin{figure}[!h]
\begin{center}
	\includegraphics[width=0.49\textwidth]{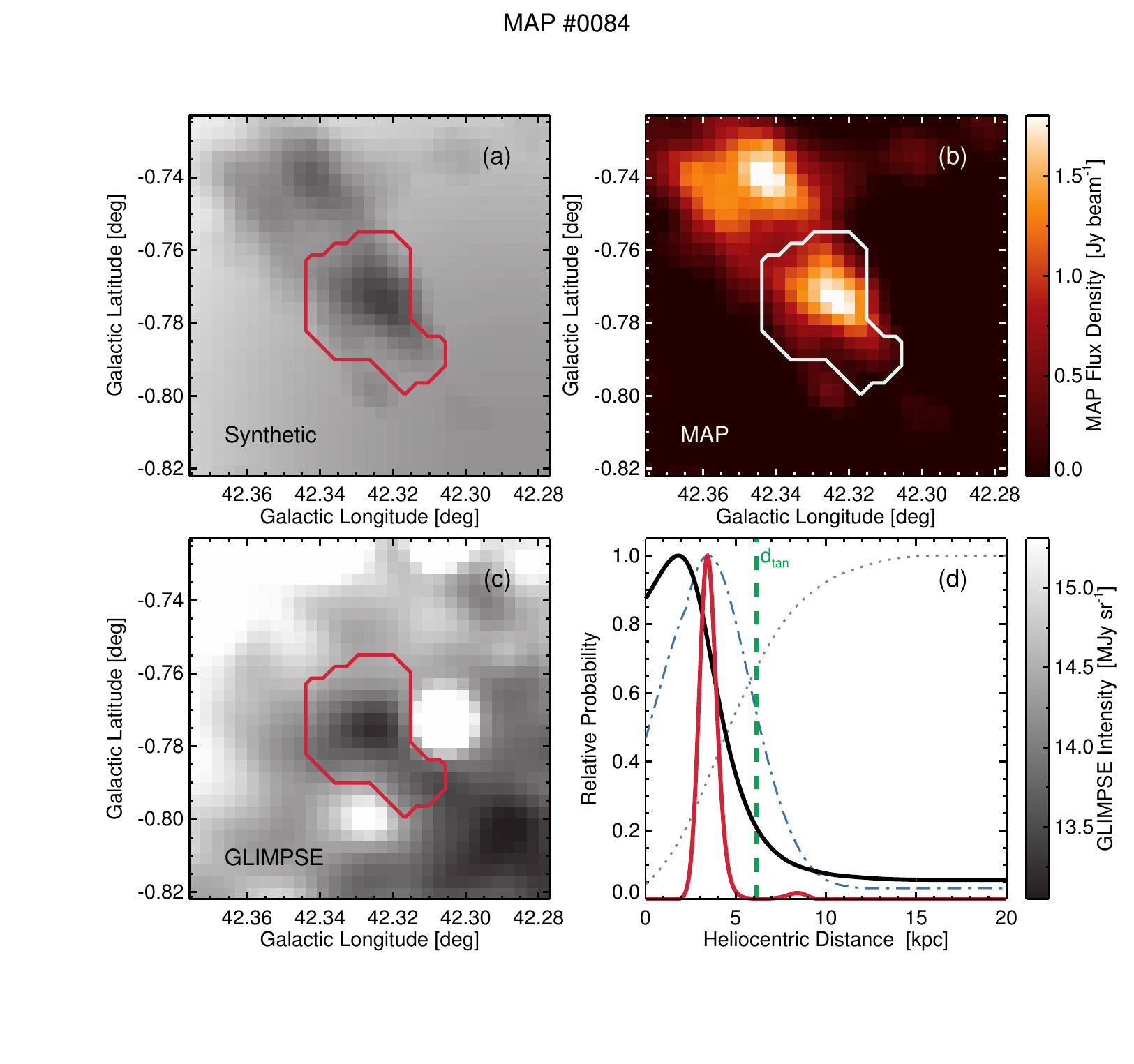}
	\includegraphics[width=0.49\textwidth]{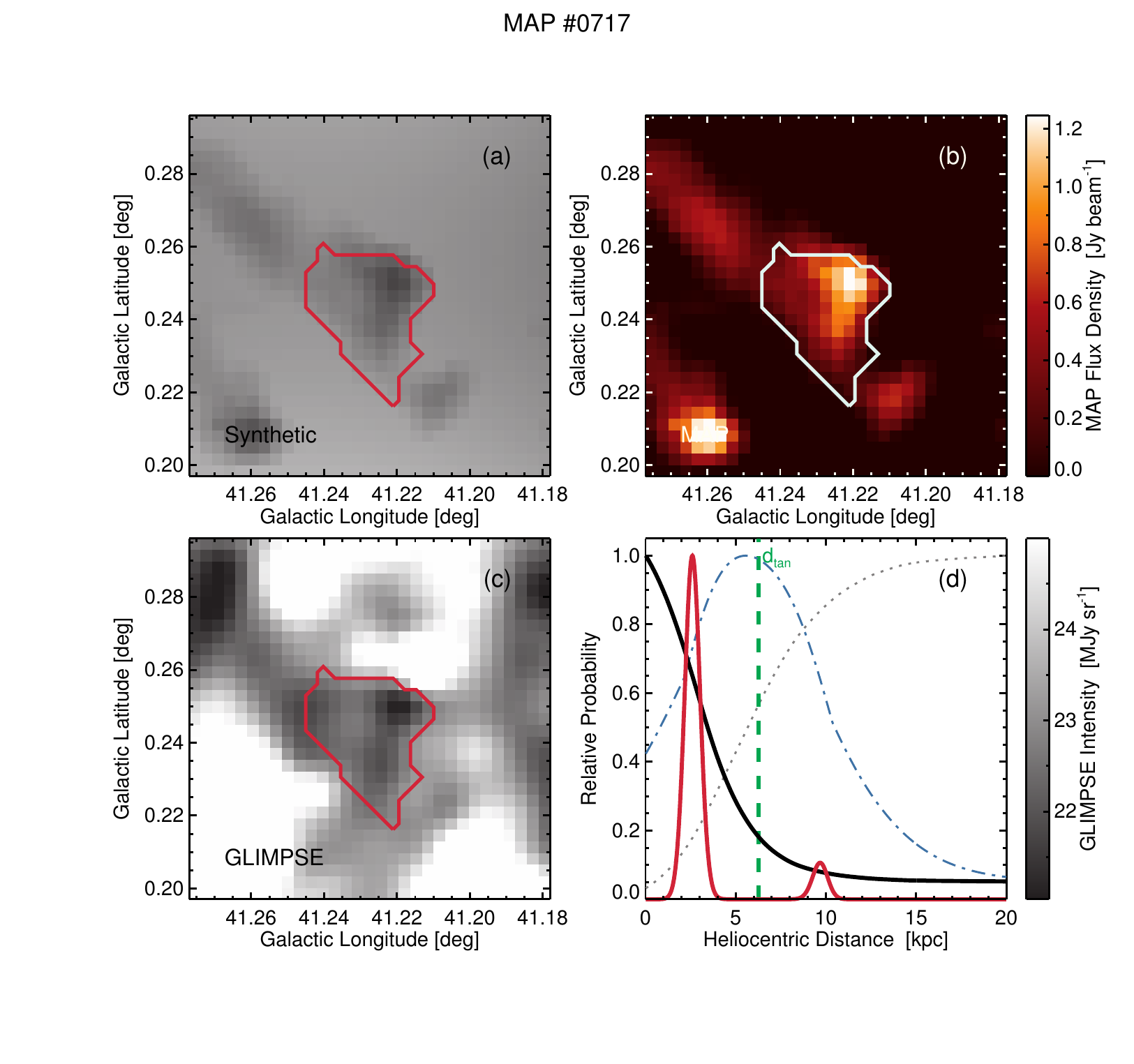}
\caption{Two examples of EMAF morphological matching in the $\ell=41^\circ$ region. Panels (a) on the left and right show the synthetic GLIMPSE map based on the SPIRE 500 $\mu$m map shown in the (b) panels. The synthetic maps shown are the best match to the GLIMPSE map shown in panels (c). Pixels with 8 $\mu$m emission instead of absorption were excluded from the morphological matching. DPDFs are shown in panels (d), with black and red corresponding to EMAF and posterior DPDFs, respectively.}
\label{fig:dist}
\end{center}
\end{figure}

Three additional $P_i$s were developed by \citet{BGPS8}. First, the molecular hydrogen (H$_2$) uses the small scale height of molecular gas in the disk to constrain clumps at high latitudes to be at their near kinematic distances. This is done using the three-dimensional model of the distribution of molecular gas in the disk of \citet{Wolfire03}. Second, the maser parallax $P_i$ uses distances measured by the BeSSeL survey to precisely determine distances to its 100+ cataloged sources. Third, the known distances $P_i$ associates molecular cloud clumps with giant molecular clouds (GMCs) found in the GRS catalog of \citet{Rathborne09}, and uses the GMC physical properties derived by \citet{RomanDuval10}.

\section{Results: Comparison to BGPS}

A visual comparison of Hi-GAL and BGPS using the $\ell=11^\circ$ region is seen in Figure 8. The Hi-GAL map is shown unfiltered and high-pass filtered, but the clumps were identified in the filtered map. As is readily seen, there are many more clumps found in Hi-GAL than over the same region of BGPS. Furthermore, the clumps are larger in Hi-GAL than they were found to be in BGPS. Comparing the Hi-GAL clump borders with the BGPS maps shows that much of what looked like 1/f noise in BGPS was actually sources which failed to meet the detection criteria. Hi-GAL's higher sensitivity to slightly larger angular scales allows us to identify this flux as coming from clumps with greater confidence than was possible with BGPS. Conversely, BGPS identified almost nothing which was not identified in Hi-GAL, confirming the low false positive rate of BGPS, which was previously derived from simulations.

\begin{figure}[!h]
		\begin{center}
		\includegraphics[width=0.49\textwidth] {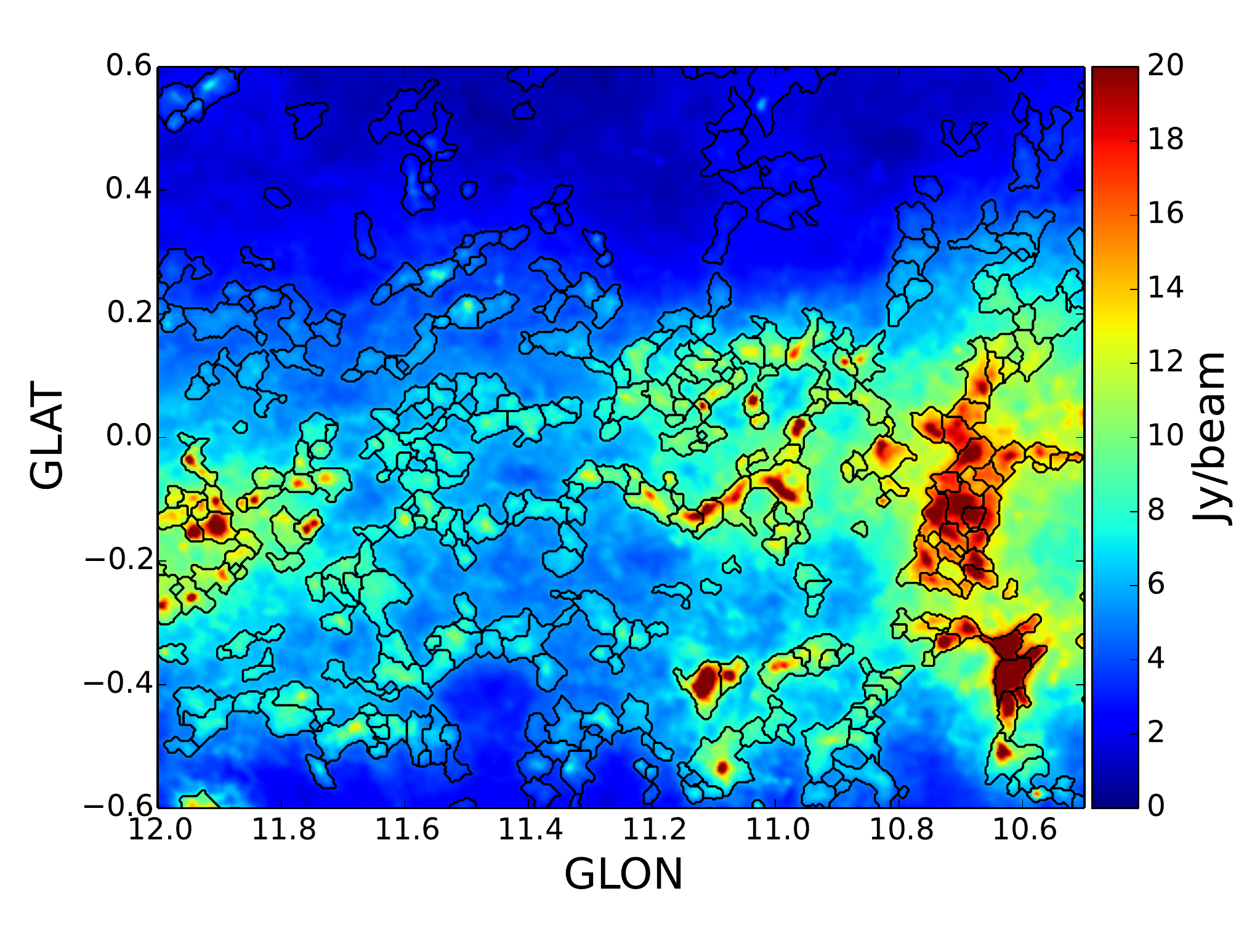}
		\includegraphics[width=0.50\textwidth] {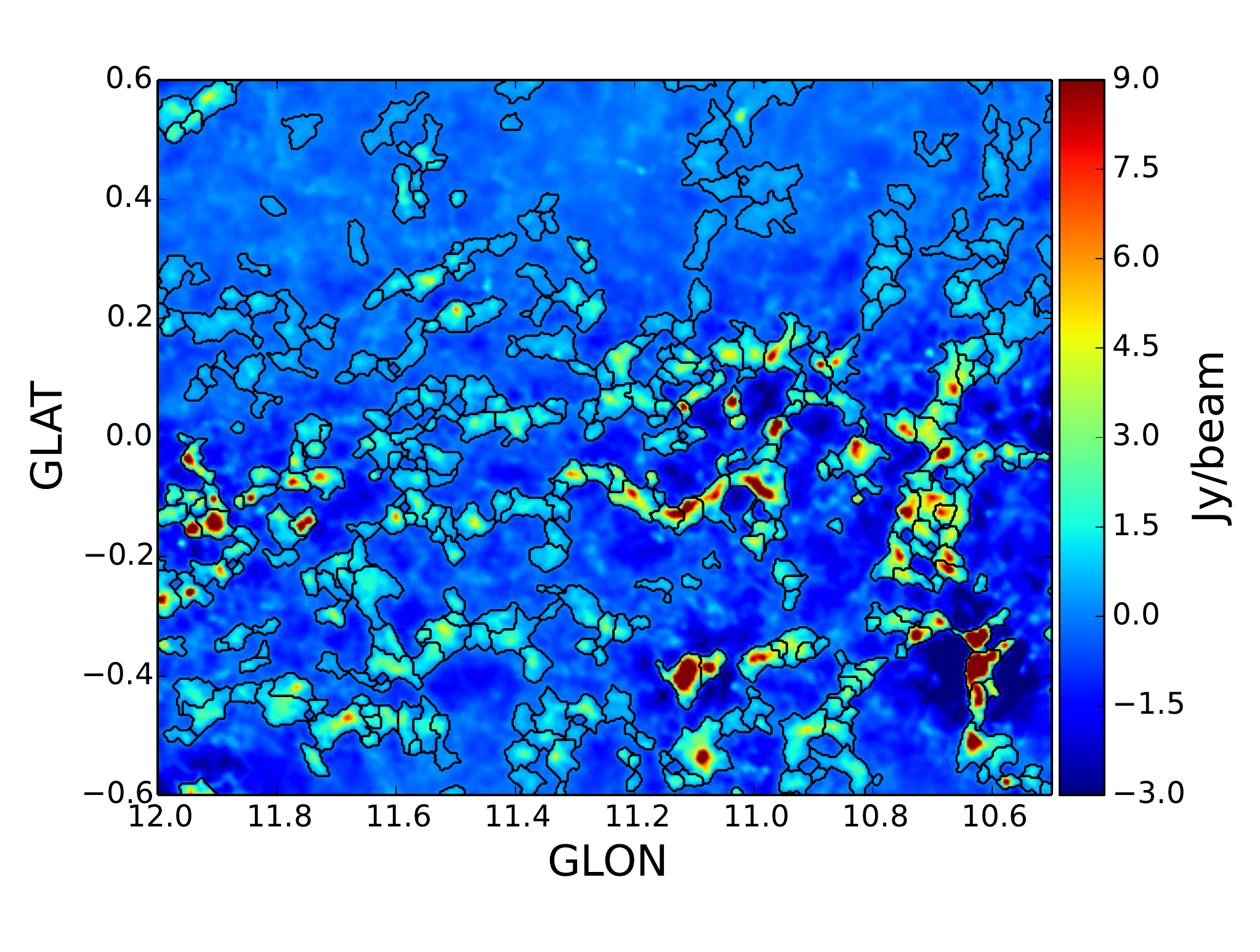}
		\includegraphics[width=0.49\textwidth] {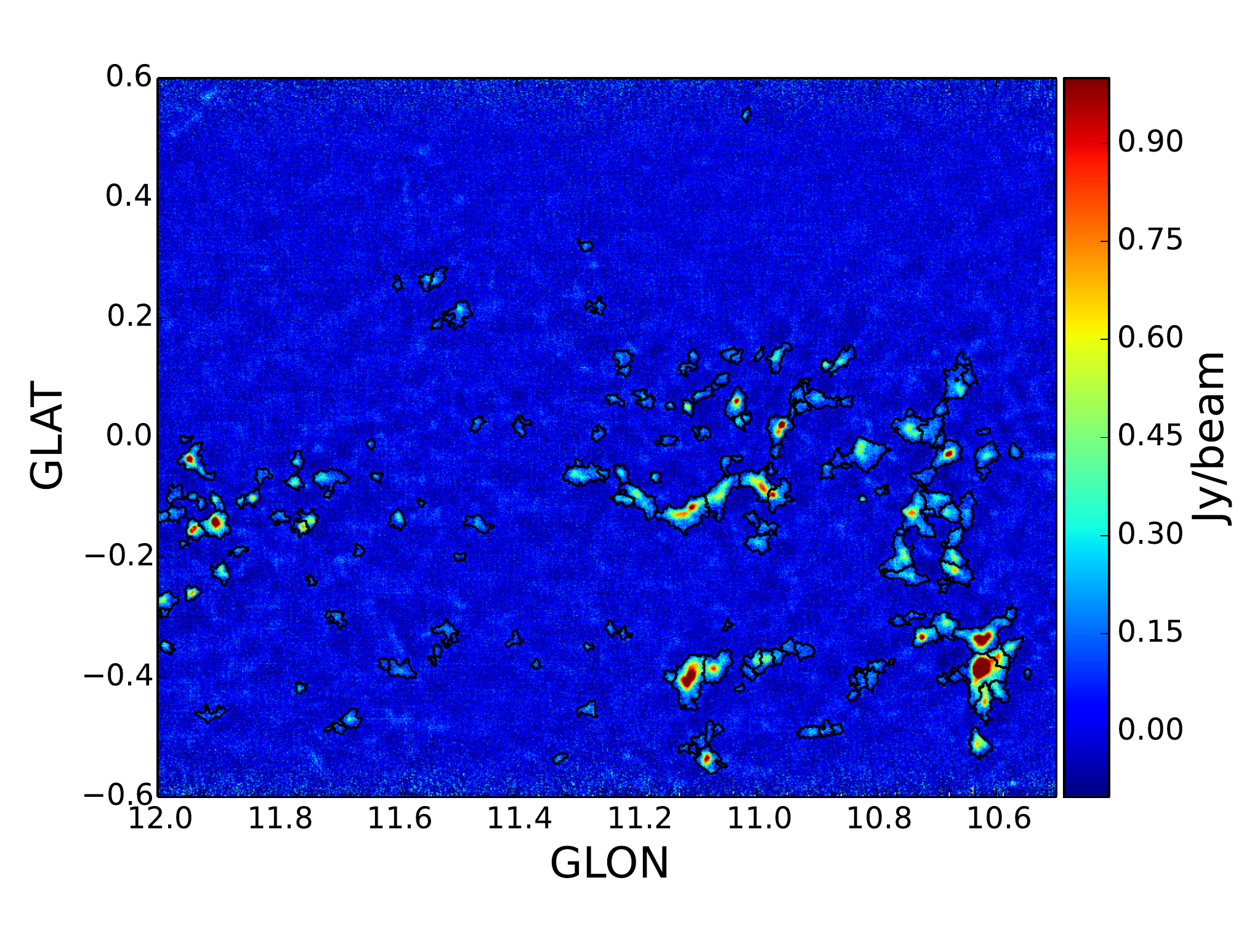}
		\end{center}
	\caption{Comparison of molecular cloud clumps identified in the same $\ell=11^\circ$ region of Hi-GAL 500 $\mu$m (\textit{upper}: unfiltered \textit{left}, high-pass filtered \textit{right}) and BGPS (\textit{lower}). Hi-GAL's higher S/N and absence of atmospheric emission allow for more clumps to be identified than in the ground-based BGPS.}
	\label{fig:l11comp}
\end{figure}

\subsection{Clump Densities on the Sky}
Table 1 shows the number of clumps in a selection of complete Hi-GAL regions, as well as all of BGPS. Each region is $2^\circ \times 2^\circ$. Listed are the number of total clumps, the number of clumps with some distance information, and the number of clumps with well-constrained distances. Clumps were considered to have ``well-constrained" distances if their DPDFs had a $\text{FW}_{68} \le 2.3$ kpc. That is, the isoprobability confidence region surrounding the distance of maximum likelihood which contains 68.3\% of the total integrated probability has a full width less than 2.3 kpc. This was adopted from \citet{BGPS8}, which found an empirical cut-off at this point such that clumps with $\text{FW}_{68} \le 2.3$ kpc either had 78\% of the total integrated probability within the most likely kinematic peak, or were located at the tangent distance. Clumps with kinematic distances, regardless of whether the posterior DPDF met the well-constrained criteria, are considered to have some distance information. GRS observations are not available in $\ell=11^\circ,202^\circ, \text{or } 217^\circ$, but there are directed observations of dense gas tracers in $\ell=11^\circ$. There is a slight tendency towards more clumps being identified in regions with lower levels of confusion, such as those in the outer galaxy.

\begin{deluxetable}{l c c c}
\tabletypesize{\scriptsize}
\tablewidth{0pc}
\tablecaption{Distance Statistics}
\tablecolumns{4}
\label{tab:Hnums}
\tablehead{
\colhead{Region\tablenotemark{a}}
& \colhead{Total}
& \colhead{Some distance}
& \colhead{Well-constrained}
\\
\colhead{}
& \colhead{clumps}
& \colhead{information}
& \colhead{distance}
}
\startdata
$\ell=11^\circ$  & 875        & 96         & 40                          \\
$\ell=30^\circ$  & 979        & 588       & 320                         \\
$\ell=41^\circ$  & 1027      & 567       & 243                         \\
$\ell=50^\circ$  & 923        & 582       & 392                         \\
$\ell=202^\circ$ & 1346     & 0           & 0                           \\
$\ell=217^\circ$ & 1048     & 0           & 0                           \\
sum                   & 6198     & 1833     & 995                         \\ \midrule
BGPS\tablenotemark{b}                & 8594     & 3508     & 1710                        \\
\enddata
\tablenotetext{a}{Each region is Hi-GAL unless otherwise specified, with Hi-GAL regions having dimensions of $2^\circ \times 2^\circ$.}
\tablenotetext{b}{Data are for the entirety of BGPS.}
\end{deluxetable}

Properties of clumps individually could only be compared where clumps identified in each survey aligned with one another. Therefore the overlapping subsections of each map pair were studied for such clumps. Table 2 compares the number of clumps found in these overlapping Hi-GAL and BGPS subregions. Listed are the number of total clumps, the number of clumps with some distance information, and the number of clumps with well-constrained distances. The reduction in number of Hi-GAL clumps in Table 2 as compared to Table 1 is primarily due to the smaller Galactic latitude range covered by BGPS as compared to Hi-GAL. Note that the regions are not consistently sized, and thus comparing numbers between regions is not a useful exercise. Information on the individual clumps being compared is listed in the Appendix.

\begin{deluxetable}{l c l c c c}
\tabletypesize{\scriptsize}
\tablewidth{0pc}
\tablecaption{Distance Statistics: Overlapping Subregions}
\tablecolumns{6}
\label{tab:regcomp}
\tablehead{
\colhead{Region}
& \colhead{Area}
& \colhead{Survey}
& \colhead{Total}
& \colhead{Some distance}
& \colhead{Well-constrained}
\\
\colhead{}
& \colhead{(deg$^2$)}
& \colhead{}
& \colhead{clumps}
& \colhead{information}
& \colhead{distance}
}
\startdata
\multirow{2}{*}{$\ell=11^\circ$} & \multirow{2}{*}{1.80}  & Hi-GAL & 376           & 53        & 30      \\
									   && BGPS   & 190           & 57        & 34      \\
&&&&&\\
\multirow{2}{*}{$\ell=30^\circ$} & \multirow{2}{*}{1.44}  & Hi-GAL & 394           & 234       & 95      \\
									    && BGPS   & 238           & 187       & 73      \\
&&&&&\\
\multirow{2}{*}{$\ell=41^\circ$} & \multirow{2}{*}{2.64}  & Hi-GAL & 650           & 383       & 107     \\
									   && BGPS   & 63            & 55        & 20      \\
&&&&&\\
\multirow{2}{*}{$\ell=50^\circ$} & \multirow{2}{*}{2.64}  & Hi-GAL & 575           & 385       & 232     \\
									   && BGPS   & 57            & 48        & 33      \\
&&&&&\\
\multirow{2}{*}{$\ell=202^\circ$} & \multirow{2}{*}{0.85} & Hi-GAL & 171           & 0         & 0 \\
									    && BGPS   & 10            & 0         & 0       \\ 
&&&&&\\
\multirow{2}{*}{$\ell=217^\circ$} & \multirow{2}{*}{1.89} & Hi-GAL & 323           & 0         & 0 \\
									    && BGPS   & 15            & 0         & 0       \\ 
\enddata
\end{deluxetable}

In the $\ell=11^\circ$ region, the numbers would indicate more clumps with distance information in BGPS than Hi-GAL. However, upon examination of the maps, it is apparent that this is only due to 5 single Hi-GAL clumps being split into two BGPS clumps each, and one single BGPS clump being split into two Hi-GAL clumps. Thus the numbers of clumps with distance information are equal in this region. There are no additional clumps with distance information in Hi-GAL for this region due to it not being covered by GRS. The only velocities are thus from directed observations of dense gas tracers done for BGPS. All additional distances are due to GRS. While the known distances prior, which associates nearby clumps with one another, was used, it did not provide any additional distances in this subset of Hi-GAL observations. The $\ell=202^\circ$ and $217^\circ$ regions has no distance information due to the lack of GRS observations in the outer Galaxy and a lack of directed observations.

\subsection{Angular Sizes} \label{sec:sizes}
The distributions of deconvolved angular radii is shown in Figure 9 for Hi-GAL (left) and BGPS (right). \citet{BGPS2} defined the deconvolved radius $\theta_R$ of a clump as the geometric mean of the deconvolved major and minor axes of the flux density distribution,
\begin{equation}
\theta_R = \eta [(\sigma_{\text{maj}}^2 - \sigma_{\text{beam}}^2)(\sigma_{\text{min}}^2 - \sigma_{\text{beam}}^2)]^{1/4}.
\label{eq:dc}
\end{equation}
For Hi-GAL $\sigma_{\text{beam}} = \theta_{\text{FWHM}}/\sqrt{8\ln{2}} = 15''$, $\theta_{\text{FWHM}} = 35''$, and $\eta = 2.4$ is a factor relating the rms size of the emission distribution to the true size of the source, which is adopted from \citet{BGPS2}. Note that if $\sigma_{\text{min}} < \sigma_{\text{beam}}$ the  deconvolved angular radius is non-real. This is the case for 14\% of our sources, arising from finite S/N. Mean angular radii for Hi-GAL and BGPS sources are $71 \pm 33\arcsec$ and $51 \pm 24\arcsec$, respectively. Both follow log-normal distributions (with shape parameters of $\sigma = 0.34\pm0.02$ and $\sigma = 0.37\pm0.01$, respectively), shown in red. We expect that the lognormal distribution is a result of random processes in the observations (such as variations in clump distances and intrinsic sizes), an interpretation which is consistent with the central limit theorem. Thus, Figure 9 is intended to demonstrate the difference in typical sizes found in the two surveys.

\begin{figure}[!h]
		\includegraphics[width=0.5\textwidth] {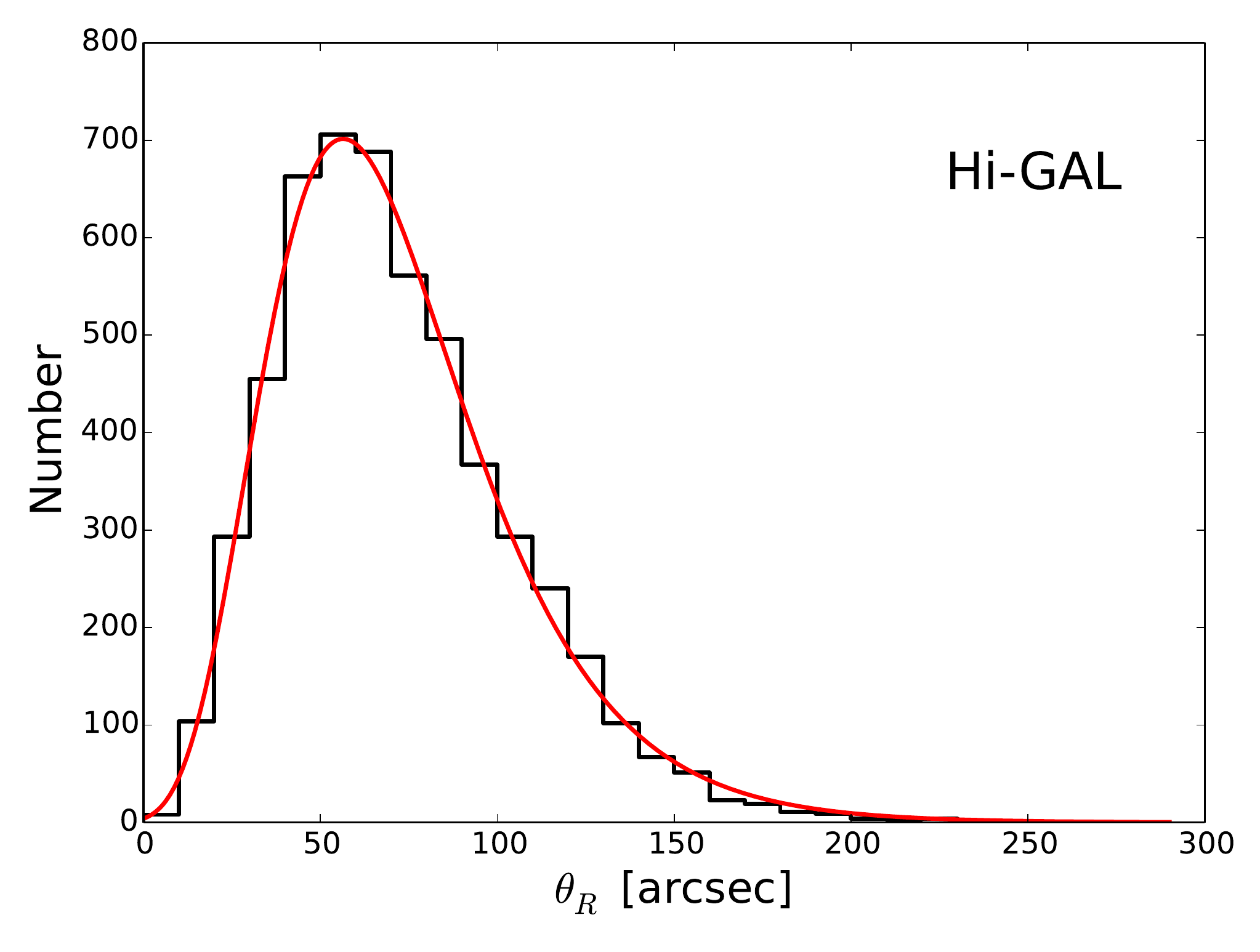}
		\includegraphics[width=0.5\textwidth] {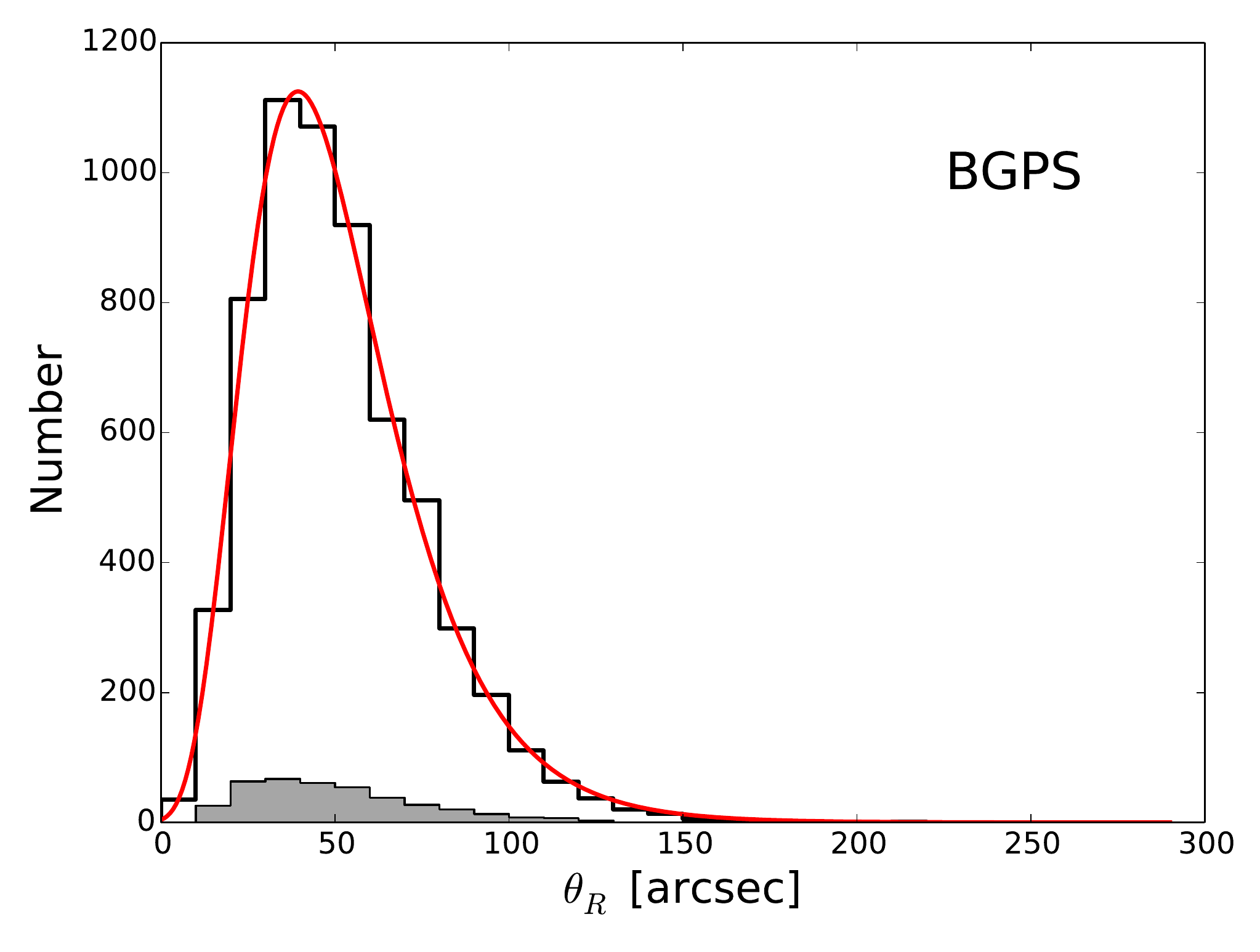}
	\caption{Distributions of deconvolved angular radii with log-normal fits in red. Hi-GAL is shown on the left, including clumps from all of the regions in this substudy, and has a shape parameter of $\sigma = 0.34$. BGPS is shown on the right (with all resolved sources included) and has a shape parameter of $\sigma = 0.37$. In grey is shown the subsample of BGPS clumps found in regions overlapping our Hi-GAL substudy.}
	\label{fig:aradii}
\end{figure}


A comparison of the angular radii of clumps matched between Hi-GAL and BGPS is seen in Figure 10. While we do not expect a consistent one-to-one ratio in angular radii, we do expect Hi-GAL clumps to be larger, and this is generally what we see, although less so with those clumps found in the $\ell=30^\circ$ region. Dashed lines show a factor of 2 in slope on either side of the identity line and demonstrate the dearth of objects in the lower right of the plot, as compared to the upper left. Clumps can be significantly larger in Hi-GAL due to the dimmer edges of the clumps being above the noise in Hi-GAL, where they were not in BGPS. The $\ell=30^\circ$ region contains the edge of the Galactic bar, the bright emission and high confusion levels of which led to the high-pass filter creating more extensive negative bowls. This subtracted more flux than was ideal around the bright Galactic bar, and thus removed flux from the comparatively dim edges of the clumps in this region, to the extent that they were found to be smaller in Hi-GAL than in BGPS. Also of note are the clumps on the left edge of this plot. These are clumps which were unresolved in BGPS, but which we can now resolve with Hi-GAL because Hi-GAL detected faint, extended emission.

\begin{figure}[!h]
\begin{center}
	\includegraphics[width=0.425\textwidth]{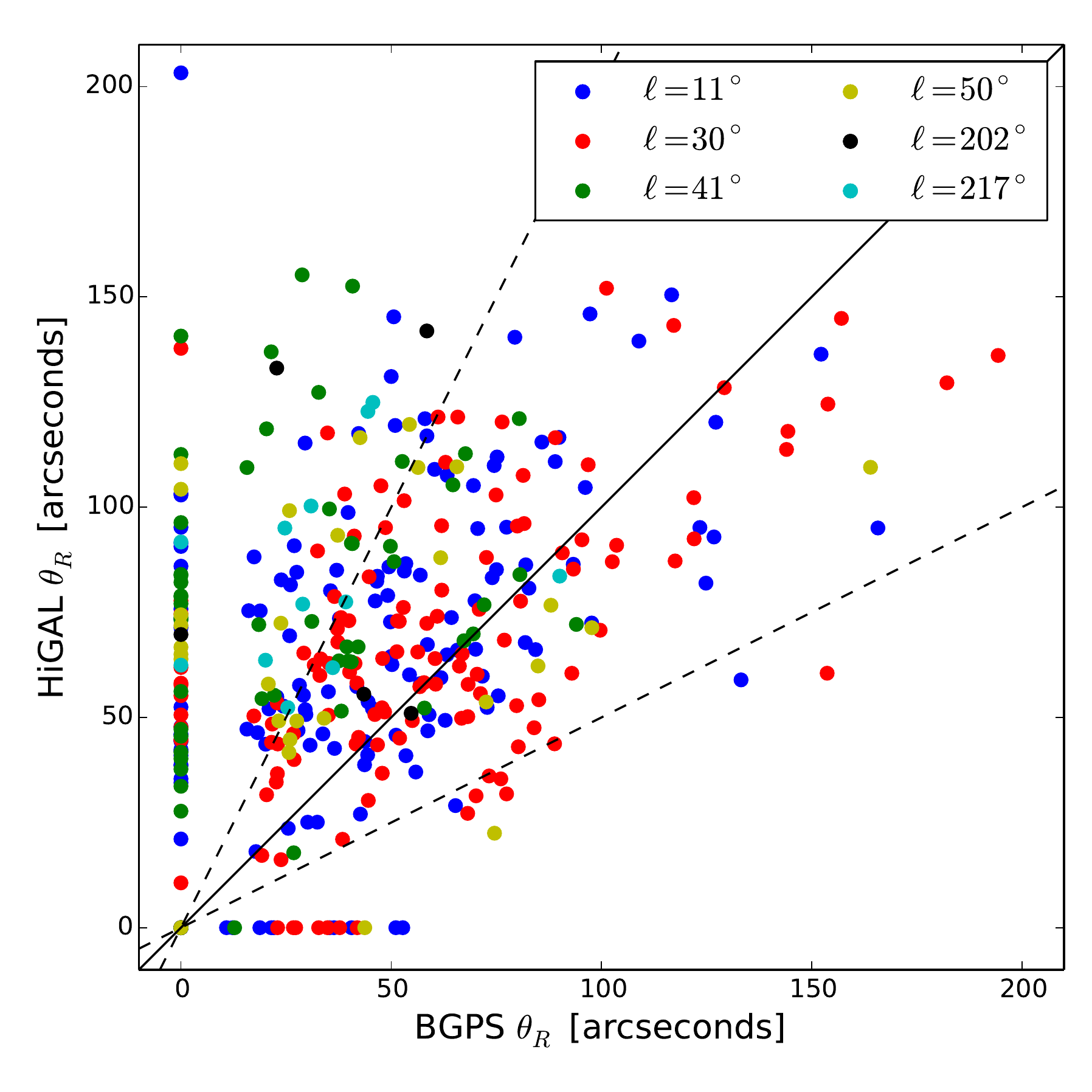}
	\includegraphics[width=0.565\textwidth]{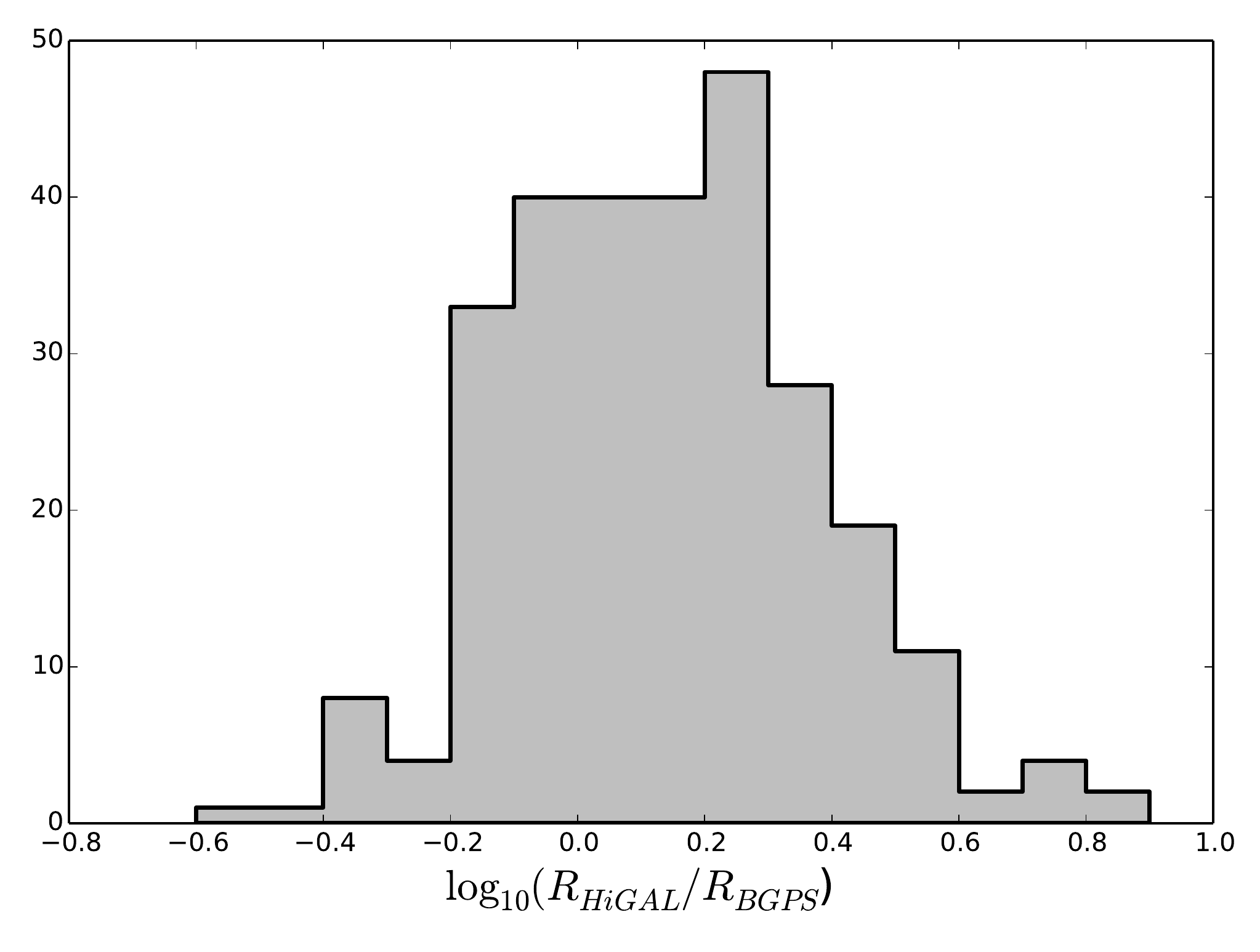}
\caption{\textit{Left}: Angular radius in Hi-GAL plotted against angular radius in BGPS, for clumps found in both surveys. Only those clumps which had well constrained distances in both surveys were used for this comparison. Colors correspond to the Hi-GAL region in which the clump was found. Note that in the line of unresolved BGPS clumps on the left of the plot, sources from the $\ell=11^\circ$ and $30^\circ$ regions are covered by later regions, and not lacking. The solid line is where the radii are equal, while the dashed lines show a factor of two in either direction. \textit{Right}: Hi-GAL to BGPS radius ratios of matching clumps, excluding those which are unresolved. The mean $\log_{10}(R_\text{HiGAL}/R_\text{BGPS})$ is 0.14, with a standard deviation of 0.24.}
\label{fig:ar_scatter}
\end{center}
\end{figure}

Figure 10 also shows the ratio of radii in the two surveys, for the same objects, but with unresolved objects excluded. The median and mean radius ratios are 1.37 and 1.61, respectively. This agrees with the difference in overall mean radii from the two surveys, as derived from Figure 9.

\subsection{Clump Masses} \label{sec:masses}
Masses can be calculated for those clumps with well constrained distances using the high-pass filtered flux densities at 500 $\mu$m and the most probable distances from the clumps' DPDFs. The conversion from flux density to mass is 
\begin{equation}
	M = \frac{R d_\odot^2}{\kappa_{500} B_{500}(T)} S_{500},
\end{equation}
where $R=100$ is the gas-to-dust mass ratio, $\kappa=5.04$ cm$^2$g$^{-1}$ is the opacity at 500 $\mu$m \citep{OH94}, $B_{500}(T)$ is the Planck function at 500 $\mu$m, $d_\odot$ is the heliocentric distance, and $S_{500}$ is the integrated flux density. \citet{Battersby11}  used pixel-by-pixel modified blackbody fits of Hi-GAL data to determine that mid-infrared-dark molecular cloud clumps generally span the temperature range 15 K $\lesssim T \lesssim$ 25 K. Furthermore, \citet{BGPS7} found an NH$_3$ gas kinetic temperature of $\langle T_K \rangle = 17.4\pm5.5$ K for a sample of 199 BGPS sources. For consistency with BGPS, and following these findings, we assume a clump temperature of 20 K.

So as not to limit ourselves to comparing only those clumps which have well constrained distances in both surveys, we calculate mass ratios while holding heliocentric distance constant. Figure 11 shows the ratio of clump masses in Hi-GAL to masses in BGPS, using those clumps listed in the Appendix. Where the ratio is significantly greater than one, the Hi-GAL clumps have much larger angular radii than their BGPS counterparts.

\begin{figure}[!h]
\begin{center}
	\includegraphics[width=0.565\textwidth]{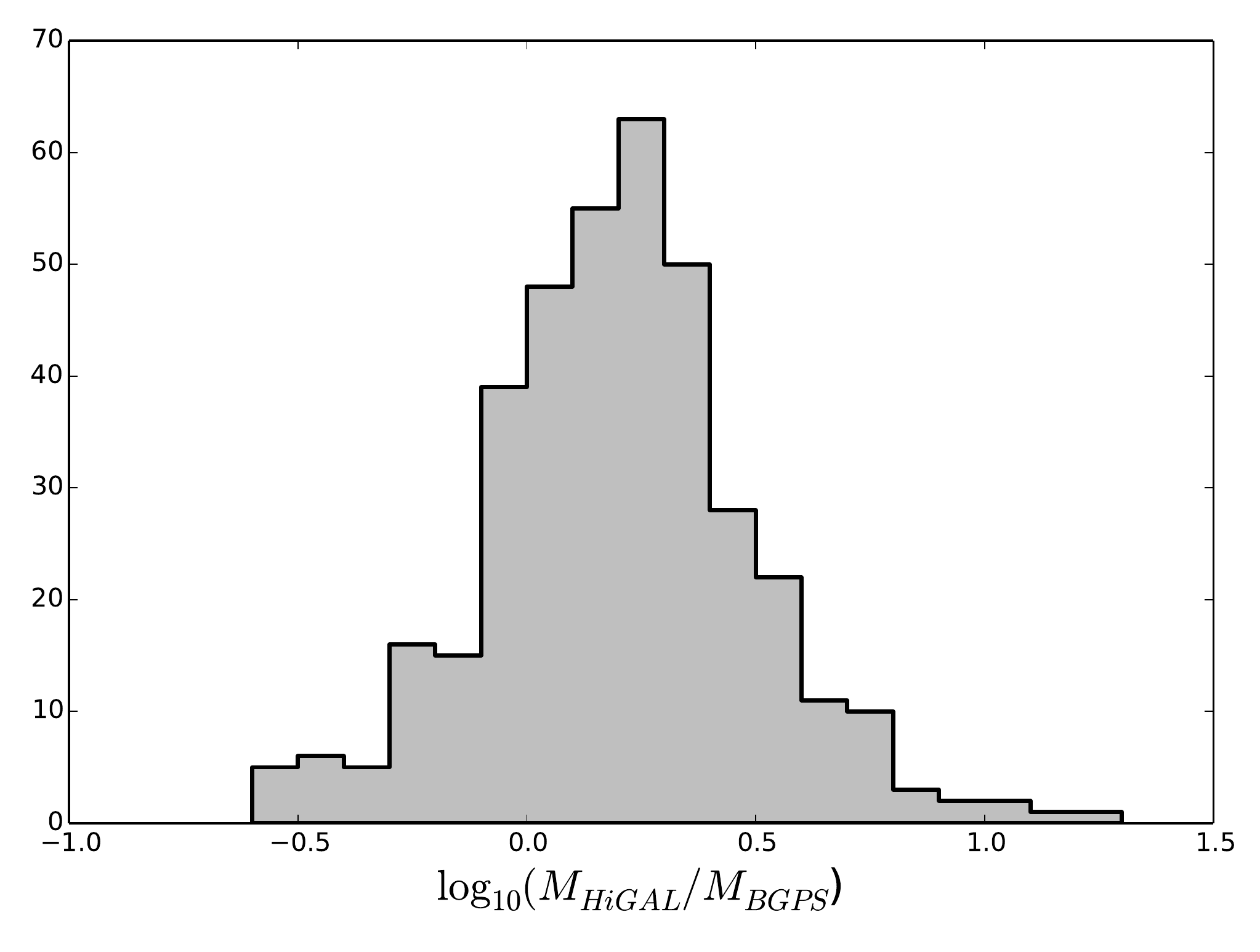}
\caption{Hi-GAL to BGPS mass ratios of all matching clumps. The mean $\log_{10}(M_\text{HiGAL}/M_\text{BGPS})$ is 0.20, with a standard deviation of 0.29.}
\label{fig:masscomp_all}
\end{center}
\end{figure}

The median mass ratio for the 384 compared clumps is 1.60, and the mean is 2.02. Taken individually, each region had a mean mass ratio within one standard deviation of the other regions. Based on the difference in mean angular radii between the two surveys found in Section 4.2, we would expect this ratio to be 2.7 under the assumption that mass scales as the cube of the angular radius. The observed mass ratio scales as the angular radius squared, suggesting that the clumps are centrally concentrated. 

With so few sources available for mass comparison, rather than the precise ratios, the meaningful results are that the median ratio of Hi-GAL mass to BGPS mass is of order unity, and that Hi-GAL masses are generally greater. The same can be said of radii, consistent with greater sensitivity to large-scale structure in Hi-GAL than BGPS, owing both to higher S/N and absence of atmospheric emission in Hi-GAL.


The distribution of masses against physical radii for all resolved Hi-GAL clumps with well-constrained distances is seen in Figure 12. Physical radii are calculated as $R = \theta_R d_\odot$, where $\theta_R$ is angular radius and $d_\odot$ is heliocentric distance. Points are colored by heliocentric distance, $d_\odot$. If mass scaled as radius cubed, as would happen if the clumps were of constant density, we would see a slope of 3. The figure shows a slope closer to 2 than to 3, thus indicating centrally concentrated masses. (The data are consistent with radial density profiles proportional to $1/R$; however, we caution against concluding a power-law profile in the absence of detailed modeling, including temperature profiles.) Malmquist bias is seen in that the smallest clumps are found at the nearest heliocentric distances, with the largest, most massive objects, being found distant to us.

\begin{figure}[!h]
\begin{center}
	\includegraphics[width=0.75\textwidth]{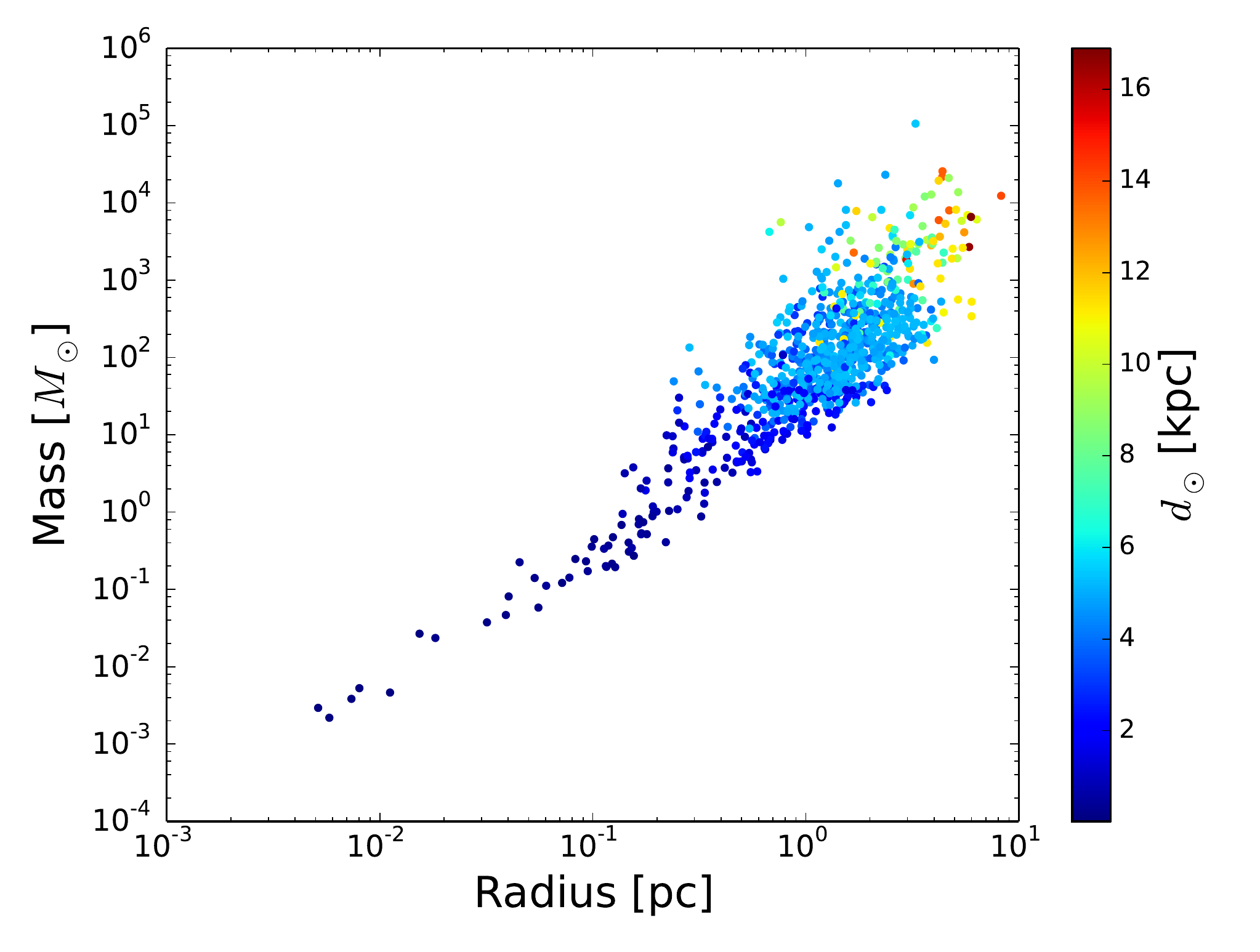}
\caption{Mass plotted against physical radius for all resolved Hi-GAL sources with well-constrained distances. Colors correspond to heliocentric distance, $d_\odot$.}
\label{fig:MR}
\end{center}
\end{figure}

\section{Discussion}


Compared to BGPS, Hi-GAL detects more sources in all Galactic regions, due to the lack of atmospheric noise. It is especially true in the outer Galaxy, where most sources which are detected with Hi-GAL were lost below the atmospheric noise in BGPS. In BGPS the number of clumps dramatically decreased as observations moved further from the Galactic center. Hi-GAL continues to see sources far from the Galactic center, as well as at Galactic latitudes farther from the the mid-plane. 

We identified a relatively constant number of clumps in each region, and, with a mean of 1033 clumps per $2^\circ \times 2^\circ$ region in our sample, we can expect nearly $2\times10^5$ clumps throughout the entire 360$^\circ$ of Hi-GAL. This is in excellent agreement with \citet{Molinari16b}, who found 85,460 clumps in Hi-GAL at 500 $\mu$m between $\ell=-70^\circ$ and $\ell=68^\circ$. When extrapolated linearly, this predicts $2.2\times10^5$ clumps in the entire Galactic plane. In those regions where GRS is available, 59\% of clumps had some distance information, and 33\% had well-constrained distances. Thus, in the full $15^\circ \le \ell \le 56^\circ$ range covered by GRS, we can expect approximately 25,000 clumps with some distance information, and 14,000 clumps with well-constrained distances. We will be including other molecular line surveys in addition to GRS: MALT90 \citep{Jackson13}, ThrUMMs (\url{http://www.astro.ufl.edu/~peterb/research/thrumms/}), and SEDIGISM (\url{http://colloques.lam.fr/GESF2014/S3/092\_SCHULLE\_Frederic.pdf}). Once these surveys and the lack of a kinematic distance ambiguity in the outer Galaxy are accounted for, we expect some distance information for 50\% of clumps and well-constrained distances for 20\% of clumps. This corresponds to 100,000 clumps and 40,000 clumps, respectively, providing a very large sample for investigating Galactic structure and physical properties of molecular cloud clumps.


BGPS used very different techniques for data reduction and large-scale structure removal than what we employed for Hi-GAL. Atmospheric emission was removed directly from the BGPS time-stream. In addition to removing atmospheric emission, this acted as an angular filter, allowing only emission on scales less than about $6\arcmin$ --- which roughly corresponded with the array field of view of $7\arcmin.5$ --- into the map. An iterative mapping algorithm then restored much of the structure removed in the sky subtraction \citep{Aguirre2011}. Since we had no need to remove atmospheric emission from Hi-GAL, we used a high-pass filter after the maps were made to remove large-scale, smooth dust emission. Despite these very different methods for large-scale structure removal, Hi-GAL and BGPS produce remarkably similar masses. The differences in masses, namely, that Hi-GAL produced generally higher masses, were expected. This is because Hi-GAL can see down to lower flux density levels and therefore can see down to the faint edges of clumps, where often BGPS could only see the brighter central regions.


Of the clumps matched between the two surveys, where the matched were one-to-one and both clumps had well-constrained distances, those distances agreed almost universally within 1 kpc. The one exception was placed at the near distance by BGPS using the HII Region Discovery Survey lookup table prior. In Hi-GAL, this prior was overridden by an EMAF not found by BGPS, which placed the clump at the far distance. Where a single clump in one survey was split in the other survey, distances for each member of the match agreed within 1 kpc, again with one exception. In this case, one Hi-GAL clumps combined two BGPS clumps, one of which had its placement at the tangent distance overridden by an EMAF placing it at the far distance.

The technique used by \citet{Molinari16b} to identify clumps in Hi-GAL was very different from our method. They used the CuTEx algorithm, which first identifies pixels above a curvature threshold. It then simultaneously fits elliptical Gaussian functions and a second-order background surface. Despite the disparate philosophies behind our clump identification techniques, our clump counts are impressively similar. Furthermore, we both see sources at all Galactic latitudes covered by Hi-GAL. Our clumps have slightly larger angular sizes, and once distances for the objects in the \citet{Molinari16b} catalog are determined, we can compare physical properties. That we found such similar results confirms that the objects we are finding are indeed clumps, thus validating both of our results.

The distribution of clumps found in our study of the $\ell=217^\circ$ region correspond well visually with what \citet{Elia13} found in their study of the same Hi-GAL region. The brighter filaments match up particularly well, and although we find more faint clumps, this is understandable, since their study was concerned only with objects for which there was a CO(1--0) velocity measurement. Although we do not yet have distances --- and therefore masses and physical radii --- for clumps in this outer Galaxy region, it can be noted that the masses and radii for our inner Galaxy clumps span a broader range than those of \citet{Elia13}. We find clumps down to smaller masses, as well as up to greater masses, and similarly for physical radii.



In future work we will extend beyond our sample of six regions and identify clumps in the entirety of Hi-GAL. We will also incorporate other molecular line surveys for the purpose of obtaining kinematic distances for clumps in areas of the Galactic plane not covered by GRS. Furthermore, we will develop two additional priors. The first relies on molecular cloud clumps' absorption of starlight at short wavelengths and uses that to determine near-infrared extinction (NIREX) distances \citep{Marshall09}. The second, H I self-absorption (HISA) and H I  emission / absorption (HIE/A) techniques, was introduced by \citet{RomanDuval09} and \citet{Anderson09}, and used on ATLASGAL clumps by \citet{Wienen15}.

\section{Conclusions}

We have made a pilot study with Hi-GAL, comparing dense molecular gas clumps in the Hi-GAL and BGPS surveys. When clumps found in both surveys are compared, derived distances agree within 1 kpc and masses are of order unity, with the mean Hi-GAL to BGPS mass ratio being 1.91. This serves as an external validation for BGPS and instills confidence as we move forward to cataloging the clumps from the entirety of Hi-GAL.

In addition to the sources which were in common with BGPS, Hi-GAL found many additional sources, primarily due to the lack of atmospheric noise. This is especially the case in the outer Galaxy. Being confusion limited, as opposed to atmospheric noise limited, means that we are able to detect clumps to lower flux densities in the outer Galaxy. Whereas BGPS produced a catalog of 8,594 clumps, 20\% of which which have well-constrained distances, we expect Hi-GAL to yield $2\times10^5$ clumps, again with 20\% having well-constrained distances.

\acknowledgements This project was supported in part by RSA 1500521 from JPL pursuant to NASA Prime Contract No. NNN12AA01C. SPIRE has been developed by a consortium of institutes led by Cardi University (UK) and including Univ. Lethbridge (Canada); NAOC (China); CEA, LAM (France); IFSI, Univ. Padua (Italy); IAC (Spain); Stockholm Observatory (Sweden); Imperial College London, RAL, UCL-MSSL, UKATC, Univ. Sussex (UK); and Caltech, JPL, NHSC, Univ. Colorado (USA). This development has been supported by national funding agencies: CSA (Canada); NAOC (China); CEA, CNES, CNRS (France); ASI (Italy); MCINN (Spain); SNSB (Sweden); STFC (UK); and NASA (USA).

ER is supported by a Discovery Grant from NSERC of Canada.

The authors wish to thank Tim Ellsworth-Bowers for helpful discussions.

Part of this work based on archival data, software or online services provided by the ASI SCIENCE DATA CENTER (ASDC).

\textit{Facility}: Herschel (SPIRE)

\bibliographystyle{apj}

\appendix

Table A1 lists the clumps matched between Hi-GAL and BGPS. Included are Galactic longitude and latitude ($\ell,b$), integrated flux density ($S$, with subscripts denoting wavelength in $\mu$m), angular radius ($\theta_R$), and heliocentric distance ($d_\odot$). Due to high-pass filtering, Hi-GAL integrated flux densities and angular radii are attenuated by $(20\pm10)\%$ and $(4\pm3)\%$, respectively, as seen in Figure 2. When a clump in one survey has multiple counterparts in the other survey, the additional counterparts are listed on the following lines.

\setcounter{table}{0}
\renewcommand{\thetable}{A\arabic{table}}



\end{document}